\documentclass[sigconf]{acmart}
\AtBeginDocument{%
  \providecommand\BibTeX{{%
    \normalfont B\kern-0.5em{\scshape i\kern-0.25em b}\kern-0.8em\TeX}}}

\copyrightyear{2022}
\acmYear{2022}
\setcopyright{rightsretained}
\acmConference[CHI '22]{CHI Conference on Human Factors in Computing Systems}{April 29-May 5, 2022}{New Orleans, LA, USA}
\acmBooktitle{CHI Conference on Human Factors in Computing Systems (CHI '22), April 29-May 5, 2022, New Orleans, LA, USA}
\acmDOI{10.1145/3491102.3502028}
\acmISBN{978-1-4503-9157-3/22/04}

\usepackage{caption}
\usepackage{subcaption}
\usepackage{multirow}

\DeclareRobustCommand{\numone}{\includegraphics[scale=0.3]{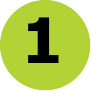}}
\DeclareRobustCommand{\numtwo}{\includegraphics[scale=0.3]{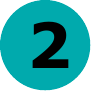}}
\DeclareRobustCommand{\numthree}{\includegraphics[scale=0.3]{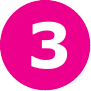}}
\DeclareRobustCommand{\numfour}{\includegraphics[scale=0.3]{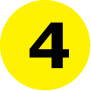}}
\DeclareRobustCommand{\numfive}{\includegraphics[scale=0.3]{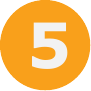}}
\DeclareRobustCommand{\numsix}{\includegraphics[scale=0.3]{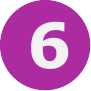}}

\begin{document}


\newcommand{\sys}{OtherTube}
\newcommand{\studyparticipantcount}{41}
\newcommand{\interviewparticipantcount}{11}
\newcommand{\studydurationindays}{10}
\newcommand{\out}[1]{{#1}}

\newcommand{\additionminor}[1]{\textcolor{black}{#1}}

\newcommand{\sang}[1]{\out{{\small\textcolor{blue}{\bf [*** Sang: #1]}}}}

\title[\sys: Exchanging YouTube Recommendations with Strangers]{\sys: Facilitating Content Discovery and Reflection by Exchanging YouTube Recommendations with Strangers}



\author{Md Momen Bhuiyan}
\affiliation{%
  \institution{Virginia Tech}
  \country{USA}
}
\email{momen@vt.edu}
\author{Carlos Augusto Bautista Isaza}
\affiliation{%
  \institution{Virginia Tech}
  \country{USA}
}
\email{carlosaugusto@vt.edu}
\author{Tanushree Mitra}
\affiliation{%
  \institution{University of Washington}
  \country{USA}
}
\email{tmitra@uw.edu}
\author{Sang Won Lee}
\affiliation{%
  \institution{Virginia Tech}
  \country{USA}
}
\email{sangwonlee@vt.edu}

\renewcommand{\shortauthors}{Md Momen Bhuiyan et al.}


\begin{abstract}
To promote engagement, recommendation algorithms on platforms like YouTube increasingly personalize users' feeds, limiting users' exposure to diverse content and depriving them of opportunities to reflect on their interests compared to others'.
In this work, we investigate how exchanging recommendations with strangers can help users discover new content and reflect.
We tested this idea by developing \sys---a browser extension for YouTube that displays strangers' personalized YouTube recommendations.
\additionminor{\sys{} allows users to (i) create an anonymized profile for social comparison, (ii) share their recommended videos  with others, and (iii) browse strangers' YouTube recommendations.}
We conducted a 10-day-long user study ($n=41$) followed by a post-study interview ($n=11$).
Our results reveal that users discovered and developed new interests from seeing \sys{} recommendations.
We identified user and content characteristics that affect interaction and engagement with exchanged recommendations; for example, younger users interacted more with \sys{}, while the perceived irrelevance of some content discouraged users from watching certain videos.
Users reflected on their interests as well as others', recognizing similarities and differences.
Our work shows promise for designs leveraging the exchange of personalized recommendations with strangers.
\end{abstract}




\maketitle

\newcommand{\rqt}[2]{\hspace{5pt} \textbf{RQ#1}. \textit{#2}}
\newcommand{\rqsub}[2]{\hspace{10pt} \textbf{#1}. \textit{#2}}

\newcommand{\qt}[2]{\begin{quote}\small{\textit{``#2''}} \hspace{5pt} - \textbf{#1}\end{quote} }

\newcommand{\inlineqt}[2]{\textit{``#2''} - \textit{#1}}

\section{Introduction}
Social media and content sharing platforms primarily use algorithms to individualize their feeds and content in order to increase user engagement~\cite{Davidson2010,fan2006personalization}. 
These algorithms typically work by predicting what users will be interested in based on their prior interaction history~\cite{covington2016deep}.
This mechanism often ends up limiting the set of content that users are likely to consume, filtering out the vast majority of content available that users could have enjoyed~\cite{Davidson2010}. 
While the resulting recommended feed may increase user engagement, users may be trapped in a ``filter bubble''---that is, isolation from alternate viewpoints---potentially limiting their choices~\cite{pariser2011filter,bessi2016users}.
However, it is challenging for users to gain awareness of their limited content consumption with all information getting intercepted by algorithms.
Users with low cognitive reflection are especially susceptible to being swayed to extreme beliefs~\cite{stecula2021social}.
Other research shows that while some people might be aware of the existence of such filters, people hardly take actions to counteract them by methods such as clearing one's browsing history, using a browser's ``incognito'' function, and clicking/liking different posts~\cite{Burbach2019}.
Though algorithmic improvement for diverse recommendations has been an active area of research~\cite{hijikata2009discovery,Wilhelm2018}, it still falls short of its goal, resulting in the persistence of filters~\cite{bryant2020youtube}.
Therefore, the limitations of modern recommender systems raise the need for design interventions that can facilitate diverse content discovery, reflection, and understanding. 

One potential intervention that could burst algorithmic bubbles is presenting diverse viewpoints to users~\cite{munson2013encouraging,ookalkar2019pop}. 
For example, in the case of YouTube, one way to implement this solution is to show users recommendations that others received; that is, a collection of videos that YouTube's algorithms recommended to other users of the platform.
We anticipate that seeing recommendations from strangers may benefit users who are otherwise exposed to a limited set of content in the following ways.
First, knowing the kinds of videos that are recommended to other users can facilitate reflection on one's own tastes and consumption behaviors through social comparison.
Prior research suggests that such comparison between peers could lead to improved self-knowledge or reflection~\cite{festinger1954theory,zhao2008identity}.
Second, seeing diverse recommendations from strangers could also facilitate the discovery of new content, such as content that simply seems interesting, content that specific groups of users watch, and content that a user did not know was available.
In addition, presenting a proxy of strangers like an anonymous persona with recommended videos can be effective for social comparison~\cite{back2010facebook,seidman2013self}.


The goal of this paper is to explore the idea of exchanging algorithmically mediated recommendations as a way to facilitate content discovery and reflection, and to assess the potential barriers to such an approach in content consumption. 
We accomplish this by designing, developing, and evaluating \sys{}---a browser plug-in for YouTube---which records the videos recommended to a user from the YouTube homepage and displays them to others.
\sys{} allows users to see strangers' YouTube recommendations (see Figure~\ref{fig:ot}) as part of the homepage.
To facilitate better social comparison, \sys{} also lets users create an anonymous persona (see Figure~\ref{fig:ot-profile}) and display it alongside their recommended videos.
Furthermore, \sys{} lets a user remove recommendations that they do not want to share.
Using our design, we aim to answer the following research questions: 

\noindent\rqt{1}{How do users discover content by browsing recommendations personalized for strangers?}

\noindent\rqt{2}{What factors affect users' interactions and engagement with recommendations personalized for strangers?}


\noindent\rqt{3}{How does browsing recommendations personalized for strangers facilitate reflection?}










To answer them, we conducted a \studydurationindays{}-day long user study with the plug-in. 
There were \studyparticipantcount{} participants in the study; participants had to use and interact with \sys{} each day and fill out a daily survey.
In addition, we logged information about the interactions that occurred within \sys{} during that period, such as the number of videos clicked and the number of clicks to see different personas.
To better understand participants' behavior during the study, we conducted semi-structured interviews with \interviewparticipantcount{} participants. 
 
Our analyses show that \sys{} can help some users, but not all, to develop new interests and rediscover old ones by seeing strangers' personalized recommendations.  
We also found that factors such as lower age and a lower need for self-reflection had a positive impact on the extent to which users browsed \sys{} content.
After viewing others' recommended videos, users understood more about their interests and how unique those interests were.
Encountering other users with similar interests also gave users a sense of belonging.
Overall, our results inform future developers of online social systems of various design considerations for allowing the exchange of recommendations: supporting content discovery and reflection needs, and encouraging cross-demographic interactions.
We conclude by exploring the implications of our approach in creating content recommendations, curating users' interests, and building social connections.

\section{Related Work}
In this section, we briefly review existing research around content discovery and reflection pertaining to recommender systems. With our system supporting comparison with strangers, we also review related literature.

\subsection{Supporting Diverse Content Discovery Online Through Recommendations}
Social recommender systems have become ubiquitous over the last decade, in areas such as social media (e.g., Facebook), e-commerce (e.g., Amazon), video sharing platforms (e.g., YouTube), and recreational services (e.g., Netflix).
As recommender systems have become highly accurate in estimating users' preferences~\cite{gunawardana2015evaluating}, \additionminor{it also comes with caveats when their suggestions become repetitive and monotonous~\cite{zhang2008avoiding}.
With the continuation of this process, it often leads to filter bubbles where algorithms discard content opposing users' interests~\cite{pariser2011filter}.}
These problems have prompted inquiry into diversifying users' exposure to differing viewpoints~\cite{resnick2013bursting}.
\additionminor{Throughout this work, we use traditional definition of exposure diversity, i.e., the supply-side (source and content) diversity~\cite{napoli2011exposure,helberger2018exposure}.}
To improve diverse exposure, one line of research takes diversification as a quality metric for recommender systems and introduces novel approaches to improve it~\cite{hijikata2009discovery,lathia2010temporal,mcnee2006being,castells2015novelty,Wilhelm2018}.
\additionminor{For example, some scholars introduced recommendations based on ``Social Diversity'', i.e, presenting recommendations from users of different social groups who are not connected~\cite{sheth2011towards}.
To generate diverse recommendations, other shcolars have suggested using content that are less popular among all users ~\cite{adomavicius2009toward}. 
Still, others have used statistical models accounting for inter-item correlation to generate diverse recommendations on YouTube~\cite{Wilhelm2018}.
While analysis of these systems deployments show some measurable improvement, they are still lacking due to the trade-off between diversity and accuracy~\cite{zhou2010solving}.
Furthermore, subsequent suggestions by recommender systems often depend on a user's prior interactions. Therefore, if a user does not interact with diverse recommendation, the following ones may not be so diverse~\cite{nguyen2014exploring}.
This issue can explain the persistent lack of diversity in most recommender system outputs, despite attempts to addressing diversity; and, our design does not depend on user interaction.
}

In parallel, there has also been some research over design-centric approach to address the issue of filter bubble on social platforms~\cite{Ookalkar2019,Medrek2018,Schaap2020,Overview78online,Gillani2018}.
These approaches include design interventions to understand users' own content consumption habit by showing information such as their topic-wise content consumption~\cite{Schaap2020}, political leaning of the sources they consume information from~\cite{munson2013encouraging}, credibility of the content they consume~\cite{bhuiyan2021nudgecred,bhuiyan2018feedreflect} and political leaning of their own social network~\cite{Gillani2018}.
Some of these approaches also promote viewpoints from alternate perspectives, such as, related content from alternate sources~\cite{Ookalkar2019} and viewpoints from a user with different political ideology~\cite{Overview78online}.
Our work adapts the approach of showing alternate viewpoints for YouTube by extending a particular demography-based feed exchanging approach to a stranger-centered feed exchanging one.

\subsection{Designing for Reflection Online Using Recommendations}
Reflection has received significant attention in HCI works.
Existing research has shown promise of reflection in various areas such as education~\cite{govaerts2012student,johnston2005amplifying,lamberty2005camera,tseng2013design}, health or wellbeing~\cite{sas2011designing,thieme2013design,gao2012design,grimes2010characteristics}, and self-knowledge or personal informatics~\cite{li2010stage,epstein2015lived,andre2011expressing,lee2011reflecting,li2011understanding}.
Several models have been proposed around personal informatics systems including Li et. al.’s Stage-Based Model of Personal Informatics Systems~\cite{li2010stage}, Epstein et al.'s the Lived Informatics Model of Personal Informatics~\cite{epstein2015lived} and Niess et. al.'s Tracker Goal Evolution Model~\cite{niess2018supporting}.
\additionminor{These models define reflection as a set of stages, from motivation to reflection~\cite{epstein2015lived}.}
Eventual purpose of many of these systems is to facilitate behavior change~\cite{consolvo2009theory,malacria2013skillometers,baumer2015reflective}, especially through goal-setting~\cite{michie2013behavior,korinek2018adaptive}.
In the past, designing for reflection around online social space has also received some attention~\cite{li2009grafitter,bae2014ripening,de2013moon}.
Some of the prior work on social media revolves around personal informatics ~\cite{li2009grafitter,de2013moon}.
For example, several works track affective expressions in users social activities (e.g., posts and comments~\cite{de2013moon,kiskola2021applying}), thus providing insight into their personal behavior.

Literature shows that many systems for reflection and desirable behavior changes---e.g., eating healthily, regular exercise, enhancing productivity---\additionminor{relies on self-tracking. However, realizing one's filter bubbles further requires elements of social comparison where such comparison can help users recognize what is missing. Literature terms this method as}
\textit{interpersonal informatics}~\cite{bales2011interpersonal}.
Works on \textit{interpersonal informatics} suggest that it can be effective for users to understand their position through projecting oneself onto others.
For example, Feustel et. al. examines reflection using cohort data from multiple sources~\cite{feustel2018people}.
Different properties of data may also stimulate reflection including showing invisible information, allowing to compare, revealing ambiguity and providing multiple perspectives~\cite{Mols2016}.
For example, some previous work shows promise of providing data for comparison~\cite{valkanova2013reveal}.
However, use of recommendations in this space is sparse.
Among the existing works, some used personalized recommendation as a tool to trigger reflection on a particular artifact~\cite{nussbaumer2012supporting,kontiza2018museum}.
Some systems also used visualization of unfiltered and curated feeds to improve users understanding of recommender systems by reflecting on them ~\cite{eslami2015feedvis,eslami2015always}.
In one sense, our work promotes interpersonal informatics in the absence of social ties, by providing data from strangers.
Though there are some HCI works using strangers in their designs, they relate to neither content discovery nor reflection~\cite{nichols2012asking,grevet2015piggyback}.
Our work draws ideas from this social comparison used for reflection and employ the approach of exchanging recommended content.


\subsection{Enabling Social Comparison for Reflection}
The theory behind behavior change leveraging social comparison is not new~\cite{festinger1954theory}. 
In some cases, such comparison could act as a support. In others, social comparison can trigger peer pressure which promotes competition~\cite{cohen1985social,ploderer2014social}.
Prior studies found this mixed effect within the same system~\cite{lin2006fish,consolvo2006design,xu2012not}.
Research also shows that constructing better self presentation for social comparison on sites like Facebook may lead to improved self concept and self-esteem~\cite{zhao2008identity,gentile2012effect}.
Motivated by these positive and negative implications of social comparison, we designed \sys{} to allow users to create an anonymous persona where they can share unidentifiable information selectively (e.g., gender, race/ethnicity and age group).
While comparison might be easier when users can identify the person they are comparing with, it also conflicts with users' need to preserve privacy for certain information~\cite{munson2013sociotechnical}.
To address this concern, Garbett and colleagues used pseudonyms and avatars, protecting users' identities in facilitating self reflection in a group setting~\cite{garbett2018thinkactive}.
For YouTube, with its long history of toxic interactions~\cite{obadimu2019identifying,chen2012detecting}, we use a similar approach to anonymize users' self-presentation (using pseudonyms, avatars and generic demographic information) in our design of \sys{}.



\begin{figure*}[t]
    \centering
    \includegraphics[width=1.\textwidth]{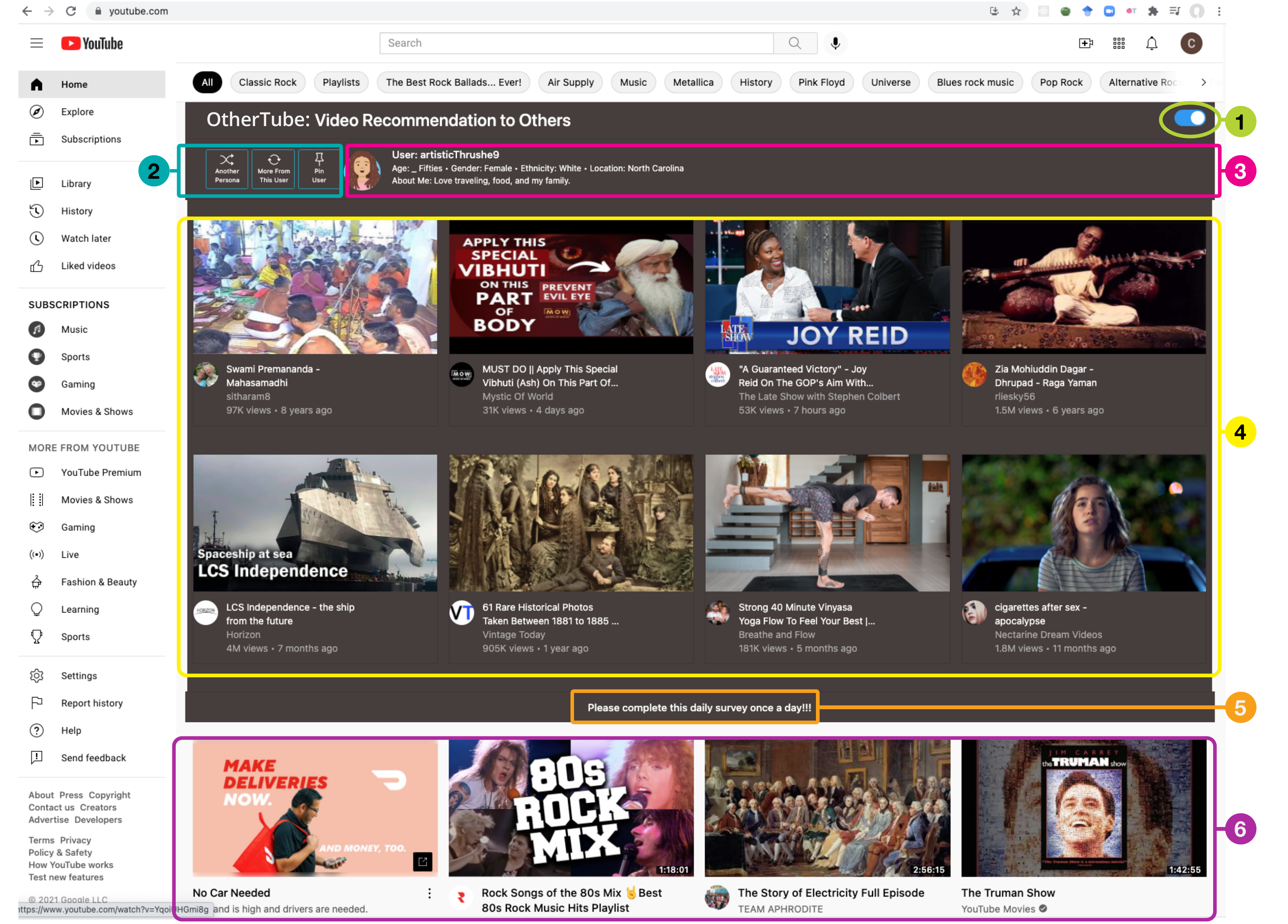}
    \caption{\sys{} embedded inside the YouTube homepage. \numone{} Show or hide the embedded content. \numtwo{} Browse different strangers or different recommendation sessions from the current stranger, and pin the current stranger. \numthree{} The stranger's profile. \numfour{} YouTube recommendations collected from this stranger. \numfive{} Link to a daily survey. \numsix{} The user's own YouTube recommendations, which \sys{} collects.}
    \label{fig:ot}
    \Description[How \sys{} is embedded in YouTube]{Once user visits their YouTube homepage, \sys{} embeds content from a random strangers' YouTube recommendations at the top of the page. This embedding looks like a container with 3 rows. First row  contains 3 buttons in the left and a profile section in the right. The next 2 rows each contains 4 videos. Below these 3 rows, there is a hyperlink to a survey.}
\end{figure*}

\section{\sys{}: Design \& Implementation}
\begin{figure*}[t]
    \centering
    \includegraphics[width=0.9\textwidth]{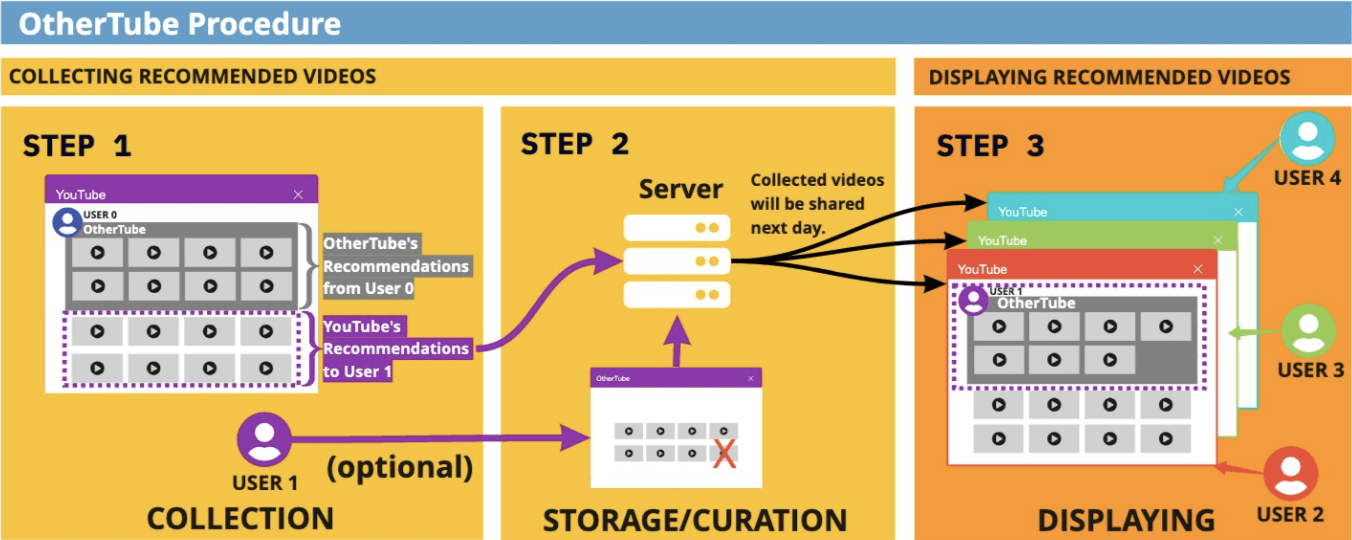}
    \vspace{-8pt}
    \caption{How \sys{} works. Each day, \sys{} collects YouTube recommendations when users access the YouTube homepage. Users have until the end of the day to remove items they do not want to share. Users can browse recommendations collected from others as recently as the previous day.}
    \vspace{-8pt}
    \label{fig:ot-how}
    \Description[Shows data collection, storage and display steps for \sys{}]{In data collection step, \sys{} extension collects YouTube recommendations to the users. Next, this data is stored in a backend server. From next day, other users can see these recommendations from the previous day. This way users have until the end of the day to remove any recommendation so that it is not shared from the next day.}
\end{figure*}

\begin{figure}
    \centering
    \includegraphics[width=0.49\textwidth]{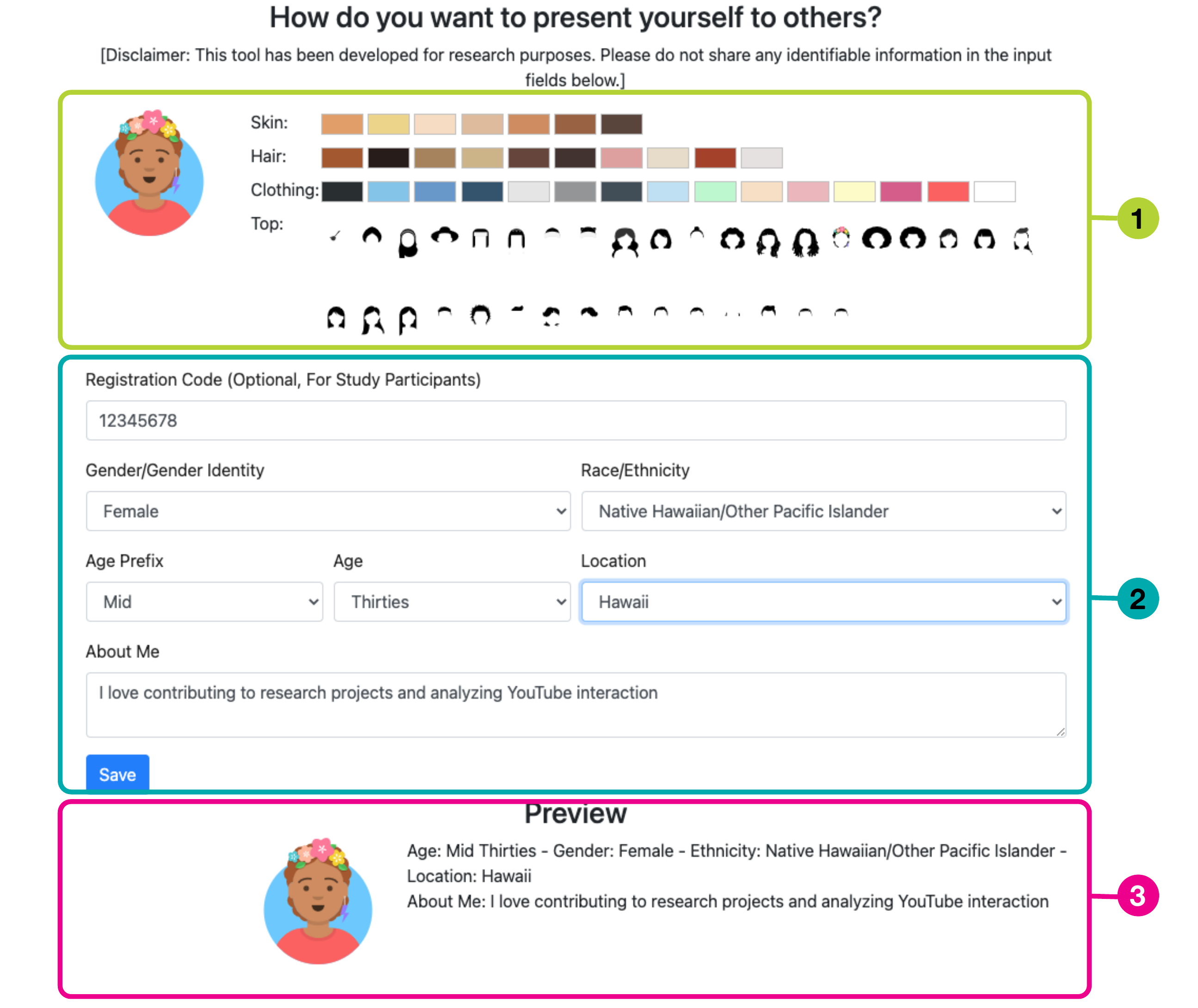}
    \vspace{-18pt}
    \captionof{figure}{\sys{}'s Options page. \numone{}~The avatar builder. \numtwo{}~Shared demographic information. \thinspace \numthree{}~How the user's profile will appear to others.}\label{fig:ot-profile}
    \Description[Page to select user profile]{User can build their profile by selecting their avatar and demographic information in this page. Avatar builder is like a selection tool for skin tone color, hair color, several types of clothing, and clothing color. Demographic information can be set by a set of dropdown menus.}
\end{figure}

\begin{figure}
    \centering
    \includegraphics[width=0.49\textwidth]{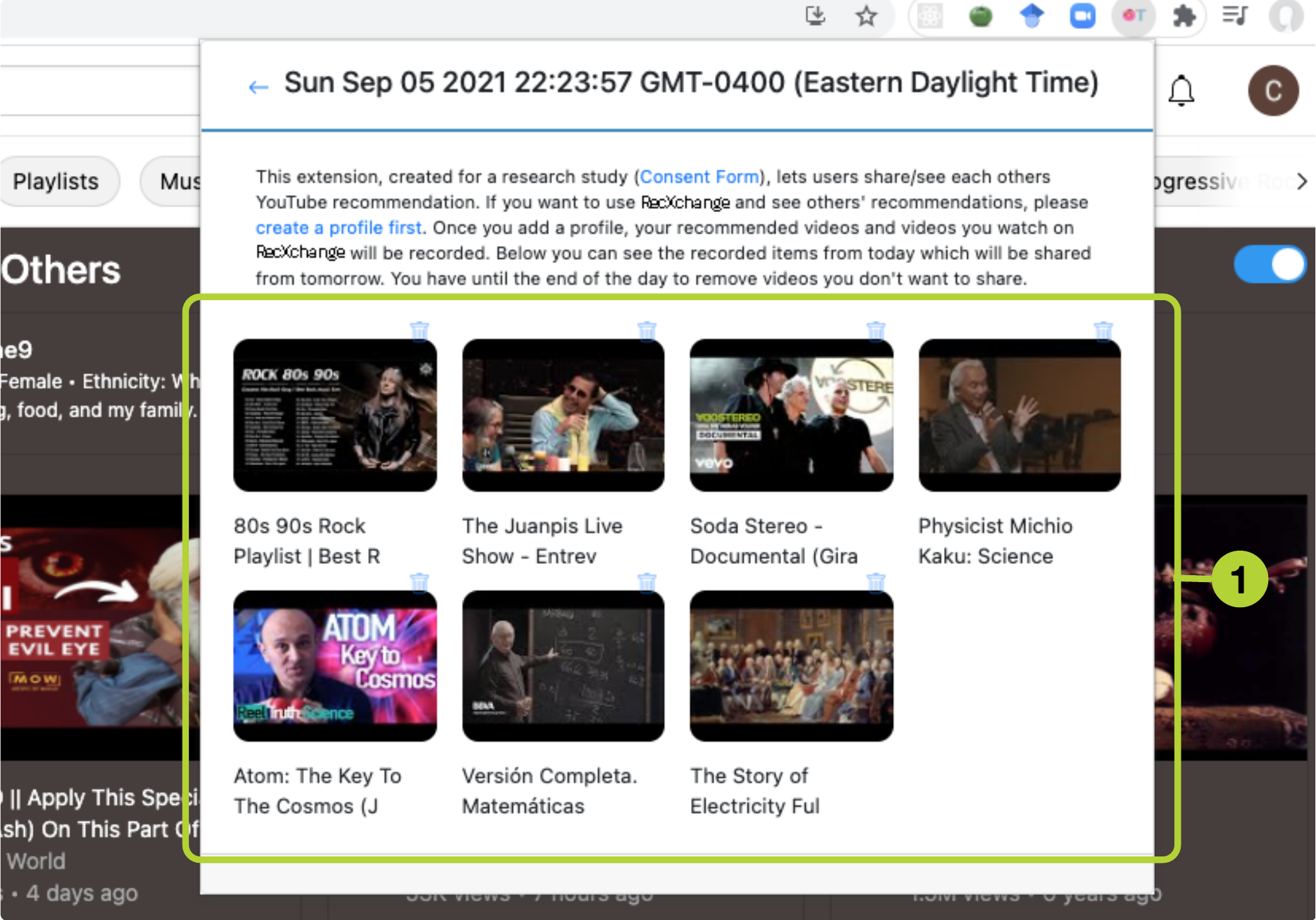}
    \caption{\sys{} Browser action page. \numone{} Collected videos with options to remove from the shared set.}\label{fig:ot-remove}
    \Description[Page showing videos collected in a session]{This page is a container with 2 rows, each row has 4 videos. At the top right corner of each video, there is a trash icon.}
\end{figure}

To provide an environment that can be integrated into users' YouTube usage, we built \sys{}. \sys{} is a Chrome extension usable across all operating systems; users need only use the Chrome browser to browse YouTube.
Our system works by collecting a user's YouTube recommendations---specifically, the top two or three rows of videos---each time a user visits the YouTube homepage.
It stores the recommended videos in a database to be shared with strangers from the next day onward. 
In short, users are given access to strangers' recommended videos (see Figure~\ref{fig:ot}-\numfour{}) in exchange for providing their own recommended videos to strangers.
Figure~\ref{fig:ot-how} demonstrates this process. Additionally, \sys{} provides three main affordances: (a) an option to allow users to create an anonymous profile (Figure~\ref{fig:ot-profile}), (b) an option to remove collected recommendations that they may not want to share (Figure~\ref{fig:ot-remove}), and (c) an option to choose between browsing strangers' profiles\footnote{Throughout the text, we use the terms \textit{persona} and \textit{profile} interchangeably.} and recommendations from users' own YouTube homepages (Figure~\ref{fig:ot}).
We describe each of these affordances below.

\subsection{Creating an Anonymous Profile}
\label{subsect:persona}
To give users extra information about the strangers whose YouTube recommendations they are browsing, \sys{} asks each user to create an anonymous profile.
For each user, we generate a random screen name---a combination of an adjective, a noun, and a number (e.g., amazedOtter4)---automatically in the back end.
Users can choose an avatar for their profile and set several demographic attributes.
Figure~\ref{fig:ot-profile} shows the available options.
The avatar builder allows users to choose a skin tone, hair color, clothing color, and appearance\footnote{We used a third-party library, \textit{AvataaarsJs}, for the avatar builder.}.
For demographic attributes, users have the option to share or not share their age, gender / gender identity, ethnicity, race, and location. 
To provide anonymity, users can only set their age as a decade-based bucket (e.g., Teen or Twenties) and location by state or province.
Users also have the option to enter their own values for certain attributes.
Apart from the demographic details, users can choose to fill out an open-ended ``About Me'' section in their profile.
For example, Figure~\ref{fig:ot} shows a profile with the text ``Love traveling, food, and my family'' in this section.
At the top of the page and the About Me input field, we put disclaimers asking users not to share any personally identifiable details.
Users can update their profiles at any time.
Finally, note that \sys{} does not start collecting recommendations from a user or showing recommendations to others until users have created a profile.
This ensures that users only see collected videos attached to profiles. 
 
\subsection{Sharing and Removing YouTube Recommendations}
\label{subsect:remove}
Each time users visit the YouTube homepage, \sys{} collects their YouTube recommendations and sends them to a back-end server, which then stores them in a database.
Going forward, we will call each visit a \textit{session}.
It is worth noting that the recommended videos are generated by YouTube's algorithm; they do not simply consist of a user's browsing history. 
This means that the collected videos do not constitute an interaction trace, so the plug-in does not have to monitor a user's entire watch history, which could be perceived as private data. 
Once a session of recommended videos is stored on the server, the recommendations are shared with other users over the following days. 
If users do not want to share certain YouTube recommendations with strangers, they can choose to remove individual videos from the set collected by \sys{}.
When users click on the extension button next to Chrome's address bar, the extension shows a list of sessions sorted by time, with the most recently collected items at the top.
While browsing the collected sessions, users can click a ``remove'' button in the upper-right corner of each video (the blue trash can icon in Figure~\ref{fig:ot-remove}) to remove content that they would rather not share with strangers.

\subsection{Browsing Strangers' YouTube Recommendations}
\label{subsect:interact}
\sys{} embeds recommendations from strangers, which are collected through the steps described in Sections~\ref{subsect:persona} and \ref{subsect:remove}, at the top of the YouTube homepage (see Figure~\ref{fig:ot}).
\additionminor{\sys{} differentiates itself from users' own YouTube recommendations on the homepage using a brown background color.}
Users can hide or show recommendations from \sys{} using a toggle in the upper-right corner of the embedded content (Figure~\ref{fig:ot}-\numone{}).
In the upper-left corner, \sys{} displays three buttons that allow users to browse strangers' recommendations (``Another Persona''), browse another recommendation session within the same stranger's persona (``More From This User''), and pin this stranger's persona (``Pin User'') (Figure~\ref{fig:ot}-\numtwo{}).
Each time users click Another Persona or More From This User, the server returns a random stranger's recommended videos or a random session from the current stranger, respectively.
When a user clicks Pin User, a shortcut to the displayed user's profile and recommended video collections is created below the button.
We added this button in case a user wants to follow and revisit a particular stranger's recommended videos. 
To the right of the three buttons, the current stranger's profile is shown, consisting of their avatar, their demographic info, and their About Me text.
Below the profile, \sys{} shows the stranger's YouTube recommendations.
Below the recommended videos, there is a link to a daily survey which we asked participants to fill out each day during our study (see Figure~\ref{fig:ot}).
Finally, \sys{} only tracks users' interactions within the plug-in (e.g., clicks on the Another Persona, More From This User, and Pin User buttons; and clicks on videos).

We built the front end of \sys{} using the React and Polymer JavaScript libraries, with Bootstrap CSS for styling.
The back end consists of a Flask-Nginx server with a MySQL database for storage.
All communication between the front end and back end is encrypted using SSL.
After building the tool, we tested it within our research groups and ran a pilot study, fixing technical issues and improving usability.
We distributed \sys{} through the Chrome Web Store.

\section{Study Deployment}
Using \sys{}, we conducted a \studydurationindays{}-day-long study.
This study was approved by the university's Institutional Review Board. 
Below, we outline our recruitment method, study procedure, data collection process, and analysis.

\subsection{Recruitment}
For our study, we aimed to recruit participants who use YouTube on a regular basis. 
In addition, we decided to recruit participants using social media, specifically using Facebook Ads, informed by others' successes in recruiting diverse populations~\cite{wang2020study,ali2020social,ramo2012broad}.
This advertising technique allowed us to reach a more diverse and targeted demography compared to Amazon Mechanical Turk or dedicated survey sites like Qualtrics~\cite{boas2020recruiting}. 
Initially, we ran an advertisement campaign targeting individuals living in the US who are 18 years of age or older, speak English, and are interested in YouTube videos.
While limiting our target demographic to those who live in the US would limit our findings, we did not want to have to account for language barriers in exchanging video recommendations. 
Our goal was to still reach diverse populations in terms of age, gender, and ethnicity. 
In addition, studying users living in a single nation provides some useful common ground upon which they can relate their interests to those of others (e.g., popular artists and domestic news items). 
The recruitment campaign was set to run for one week, from July 8, 2021 to July 15, 2021. 
We spent \$315 on the campaign and received 568 responses. 

\subsection{Procedure}
Users who clicked the Facebook advertisement were redirected to a pre-survey to sign up for the study.
At the beginning of the pre-survey, we screened users according to several criteria.
To be eligible for the study, users had to (i) be 18 or over, (ii) currently reside in the United States, (iii) visit YouTube at least once a day, (iv) typically browse YouTube on a laptop or desktop computer (as opposed to mobile-only users),  (v) typically use Chrome to browse YouTube, (vi) have English as the primary language of the YouTube content they watch, and (vii) typically start browsing YouTube from the YouTube homepage (youtube.com).
Out of the users who submitted the presurvey ($n=568$), 318 were eligible for the study.
We invited these participants via email to \additionminor{complete the consent form and} start the study\footnote{Initially, we prioritized minorities for invites to form a diverse pool. Due to the limited response, we eventually reached out to all participants.}. 
The invitations were sent out in two batches: one from July 19 through August 3, 2021, and another from August 5 through August 18, 2021. 
Note that while the emails to all participants in a given batch were sent out on the same day, users could have started using the extension on different days, resulting in batch periods exceeding 10 days.
Out of our 318 invitees, \studyparticipantcount{} participated in the study by installing the plug-in and filling out the daily survey at least once. 
We created an instructional document describing how to participate in the study. 
Users who accepted the invitation had to install the \sys{} extension from the Chrome Web Store.
After installing the extension, users had to create profiles.
At the beginning of the study, to mitigate the cold start problem, we created a research account so that participants could begin to see embedded recommendations from the first day.
For \studydurationindays{} consecutive days, users were asked to use YouTube as they normally would and interact with \sys. 
Each day, they were also asked to submit a daily survey which took about five minutes to complete.
We sent reminder emails around 6 P.M. EDT each day to remind users who had not yet submitted the daily survey.
Despite the reminders, participants did not consistently submit the survey, leaving us with 356 (8.7 on average) responses instead of 410 (41 participants × 10 days).
Upon completion of the study, we invited about half ($n=19$) of the participants to an interview based on their survey completion rates.
Of those invited, 12 participated in the interviews. 
Each meeting was recorded for analysis.
Because one user revealed that they neither followed the study instructions nor had a clear understanding of how the plug-in works, which would have left the user with no context for many of our questions, we ignored this user's response and analyzed the remaining 11 recordings.
We compensated study participants with \$25 gift cards and interview participants with \$15 gift cards, in line with federal minimum wage requirements.
 
\subsection{Participants}\label{sec:part}
\begin{figure*}[t]
    \centering
    \includegraphics[width=0.95\textwidth]{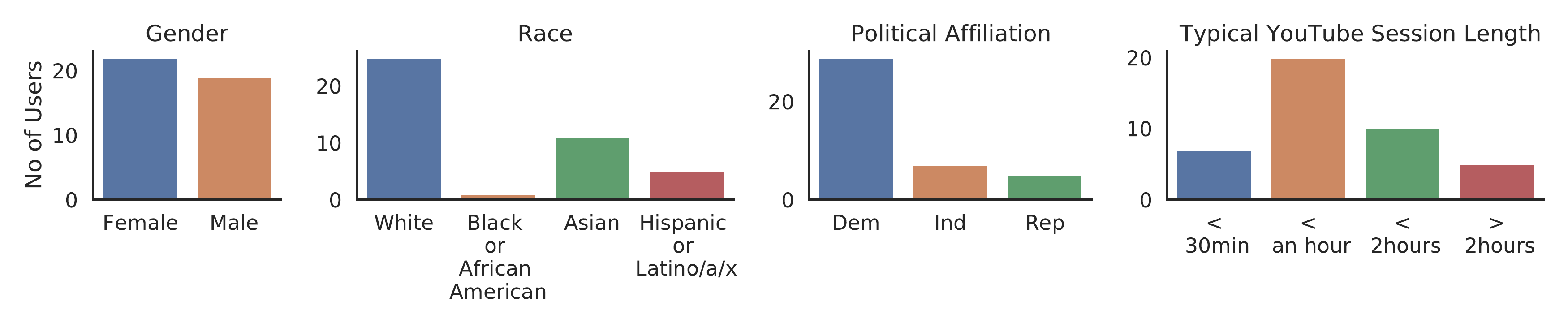}
    \vspace{-15pt}
    \caption{Demography of the participants in the study.}
    \vspace{-8pt}
    \label{fig:demo}
    \Description[Participant demography]{In this figure, there are four bar charts, respectively for gender, race, political affiliation and typical YouTube session length. Among these charts, only gender one is well-balanced.}
\end{figure*}

Our pre-study survey was mainly designed to filter out ineligible users and create a participant pool that represent diverse demographic groups.
However, we could not completely fulfill this objective due to an inconsistent response to study invites.
Figure~\ref{fig:demo} shows the distribution of age, gender, political affiliation, and length of a typical YouTube browsing session among our \studyparticipantcount{} participants.
The participants' demography is balanced in terms of gender.
However, it is heavily skewed in terms of race, with only one Black or African American participant despite a sufficient number of Black or African American users signing up for the study (see the contrast in Appendix~\ref{app:demo}).
This disparity could be caused by hesitancy towards installing tools or hesitancy towards research studies due to past injustices~\cite{koo2005challenges,hughes2017african}.
By political affiliation, the majority were Democrats.
In the About Me section of their profiles, participants mentioned various occupations (e.g., ``manager'' and ``recent college graduate'') and interests (e.g., photography, traveling, yoga, DIY, Hololive, and BTS).
 
\subsection{Data Collection}
We collected data from our participants in multiple ways, beginning with the pre-study questionnaire.
The questionnaire was followed by interaction traces and daily surveys during the study, and the post-study interviews came last.

\subsubsection{Pre-Study Questionnaire: Need for Self-Reflection and Insight}
\begin{figure*}[t]
\centering
\begin{subfigure}{.4\textwidth}
  \centering
  \includegraphics[width=.99\linewidth]{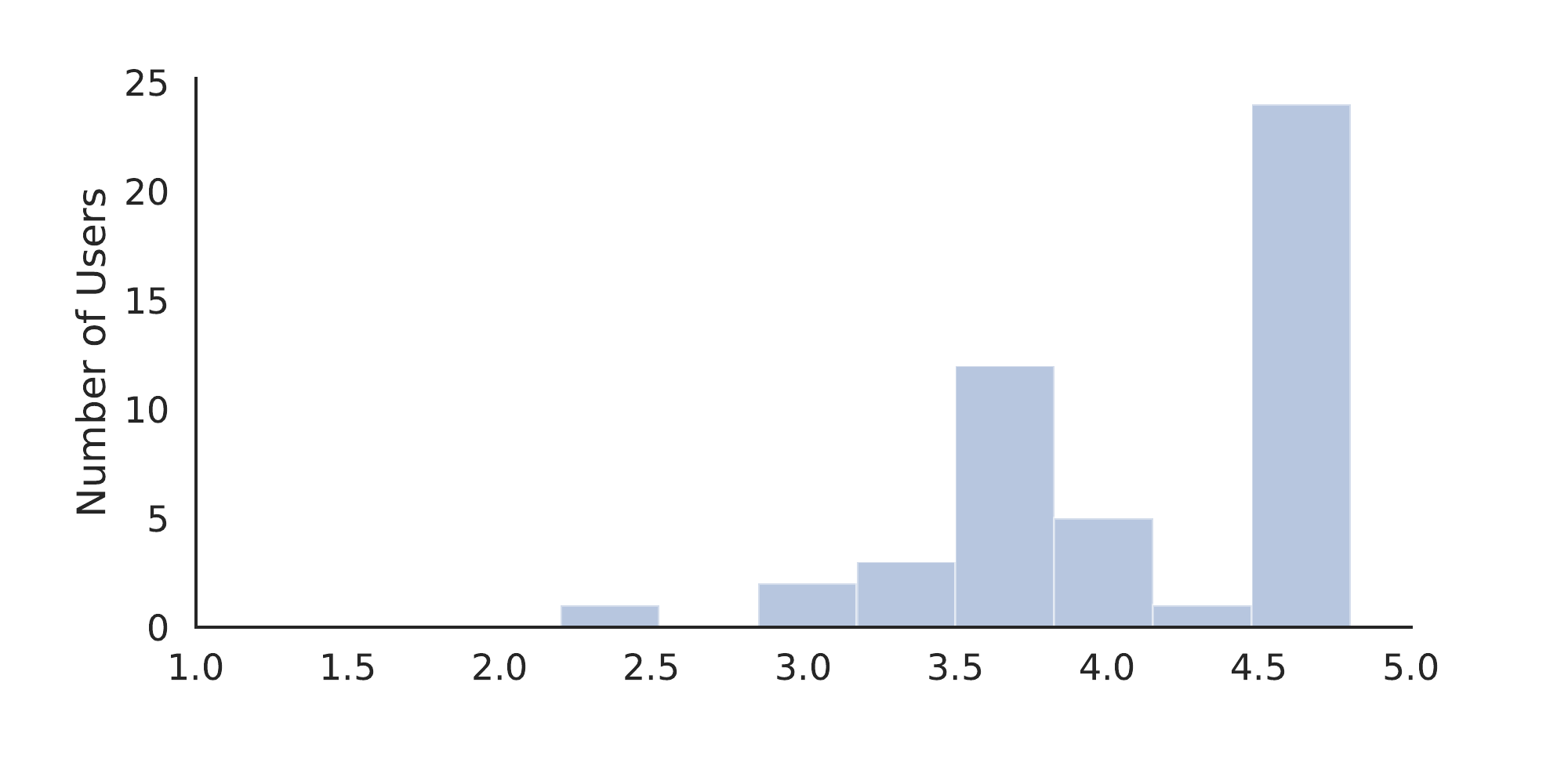}
  \vspace{-18pt}
  \caption{Need for self-reflection}
  \label{fig:nfr}
\end{subfigure}%
\begin{subfigure}{.4\textwidth}
  \centering
  \includegraphics[width=.99\linewidth]{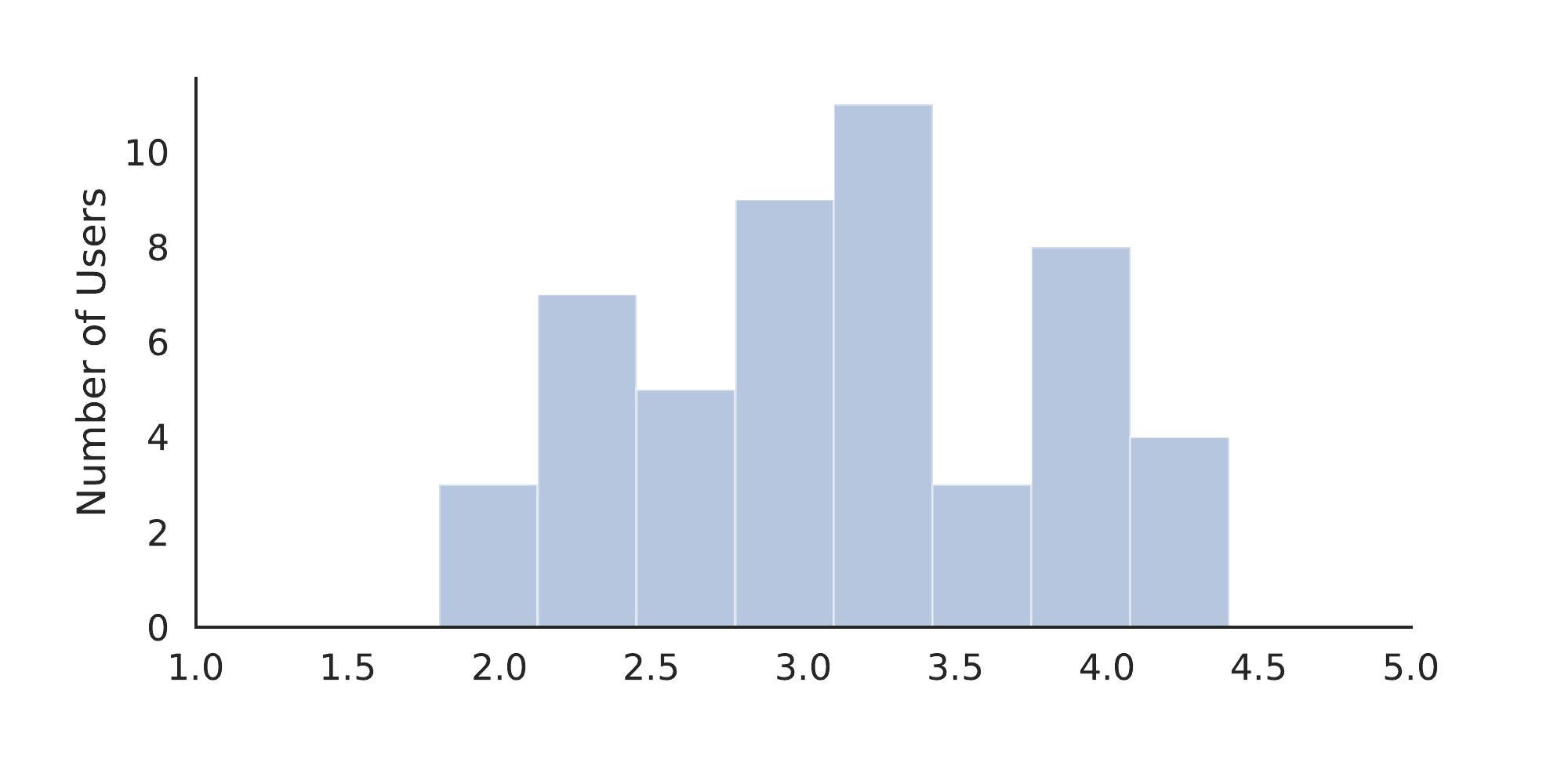}
  \vspace{-18pt}
  \caption{Insight}
  \label{fig:nfi}
\end{subfigure}
\vspace{-8pt}
\caption{Distribution of participants' need for self-reflection and insight, bucketed for ease of understanding.}
\label{fig:nfri}
\Description[Participants' responses to need for reflection and insight]{There are two bar charts in this figure. Users' need for reflection is skewed towards high where as insight looks normally distributed.}
\end{figure*}

In the pre-study questionnaire, along with demographic questions, we asked about users' need for self-reflection and insight (see Appendix~\ref{app:scale} for the items), using scales from prior research~\cite{Halttu2017} on 5-point Likert items with responses from ``strongly disagree'' (1) to ``strongly agree'' (5).
We measured these metrics to see if the self-assessed need for self-reflection and insight would correlate with how users interact with \sys{}.
Participants' responses had a high level of consistency for both questions, similar to prior studies (Cronbach $\alpha$ [Need for Self-Reflection]: 0.97, Cronbach $\alpha$ [Insight]: 0.95)~\cite{Halttu2017,Grant2002}.
For each trait, we found the mean of the items after inverting items that were phrased in the opposing sense.
Figure~\ref{fig:nfri} shows the distribution of user responses for need for self-reflection and insight.

\subsubsection{Interaction Traces from \sys{}}
We collected data on users' interactions with \sys{}, including clicks to see recommendations from different strangers, clicks to see more from a stranger, clicks to watch videos, clicks to pin strangers, and clicks to remove own recommendations.

\begin{table*}[]
    \centering
    \sffamily \scriptsize
    \begin{tabular}{p{0.08\textwidth} p{0.86\textwidth}}
        \textbf{Question Type} & \textbf{Item}\\
        \hline
        \multirow{6}{*}{Likert-scale}  & (a) Today, I saw some videos on \sys{} that caught my attention. \\
        & (b) Today, \sys{} recommended me videos that I would not have expected in my feed. \\
        & (c) Today, I discovered some new types of content on \sys{} that I would like to watch. \\
        & (d) Today, YouTube recommended me some videos that I would not have expected in my feed. \\
        & (e) Because of using \sys{}, YouTube is recommending diverse videos to me. \\
        & (f) I feel comfortable sharing my recommended videos with others on \sys{}. \\
        \hline
        \multirow{6}{*}{Open-ended} & (a) If there were any videos that caught your attention from \sys, could you tell us what they were and why you were interested in them? \\
        & (b) If you removed any recommended videos of yours from the \sys{} plug-in, could you tell us why? \\
        & (c) While using \sys, did you learn anything new about certain populations (different age group, different gender)? If so, what did you learn? \\
        & (d) While using \sys, did you learn anything new about your own taste compared to others? If so, what did you learn? \\
        & (e) What made you hesitate to watch videos from \sys, if any? \\
        \hline
    \end{tabular}
    \caption{Daily Survey Questions}
    \label{tab:daily-survey}
\end{table*}

\subsubsection{Daily Survey: Perceptions of Recommendations on \sys{}}
In the daily surveys, we asked users Likert scale and open-ended questions about their experiences with \sys{}(Table~\ref{tab:daily-survey} lists the questions). 
We included several Likert items on a 5-point scale from ``strongly disagree'' (1) to ``strongly agree'' (5) about users' perceptions of \sys{} on YouTube. 
These questions captured users' perceptions of content discovery each day, whether their own YouTube recommendations were affected by their interactions with \sys{}, and how they felt about sharing their YouTube recommendations with strangers.
Through open-ended questions, we aimed to understand participants' reflections on themselves and others, and to learn the motivation behind any actions they performed or chose not to perform (e.g., content removal or hesitation to watch content).

\subsubsection{Interview}
After completion of the \studydurationindays{}-day study, we interviewed 11 participants.
The interviews focused on understanding how they used the features of the plug-in, and whether using \sys{} encouraged content discovery or facilitated any reflection.
Additionally, after analyzing each interviewee's daily survey responses, we asked them to elaborate on any points that seemed unclear.
For example, because one user mentioned that their interests are ``conservative'' in their daily survey, we asked them to elaborate on what they meant by that (i.e., whether the user was interested in political conservatism or had conservative viewing habits with preference to watch videos with limited interests). 
We concluded the interview with some usability questions.
See Appendix~\ref{app-interview} for the complete interview questionnaire.

\subsection{Method of Analysis}
We gathered both quantitative responses (pre-study survey, Likert scale--based daily survey questions, and interaction traces) and qualitative responses (open-ended daily survey questions and interview responses). Below, we describe our analytical methodology. 

\subsubsection{Likert Items on the Daily Survey}
To gain insight into self-assessment on the statements, we checked the average responses to each Likert item on the daily survey. We also performed Mann-Whitney U tests, a nonparametric test,   
to see if using \sys{} produced any changes in their perceptions, comparing the responses from participants' first and last days of using \sys{}. 
Because not all users submitted the survey on each of the \studydurationindays{} days, the last day might not have been the \studydurationindays{}th day for each user.

\subsubsection{Interaction Traces}
To answer RQ2, we modeled the number of daily clicks on the Another Persona button using negative binomial regression\footnote{We repeated this process for the More from This User button, but the results were not significant; therefore, we omitted them from this paper.}. 
The independent variables for the regression consisted of users' demographics, the need for self-reflection, and insight. 
Because we recorded the number of clicks on each of 10 days, resulting in repeated measures for each user, we used a mixed-effects regression model. 
Additionally, because clicks are a count variable, we used negative binomial regression\footnote{Due to overdispersion, we chose a negative binomial model over Poisson regression.}.
Because some users did not click the buttons every day and random effects require multiple observations per user, we used users who clicked the button on at least three days.
Therefore, instead of 410, we had 280 data points from 32 users (an average of 8.8 data points per person) for our model.
We used \textit{mixed\_model} from the \textit{GLMMadaptive} R~package~\cite{rizopoulos2019glmmadaptive}.
For the sake of interpretation, we present marginal coefficients instead of fixed effect coefficients in this model\footnote{Given the nonlinear link function ($Log$) in our model, random effect intercepts can have a multiplicative effect, rather than additive, complicating interpretation. Therefore, following Hedeker et al.~\cite{hedeker2018note}, we extracted the marginal coefficients and their standard errors from the model using the \textit{GLMMadaptive} R~package.}. 
Additionally, we computed the Spearman rank correlation, a nonparametric statistic, to assess whether a correlation exists between users' interactions and engagement; that is, clicks on videos, Another Persona, and More from This User.
Apart from these tests, we also analyzed other simple statistics.
 
\subsubsection{Open-Ended Daily Survey Responses}
We performed thematic analysis on the open-ended daily survey responses. 
With five open-ended questions, we had 1,780 responses (356 daily survey submissions × 5 questions), 976 of which were either empty or contained unhelpful responses, like ``no'' or an incomplete response (e.g., ``Yoga videos'' for Open-ended-(a) in \ref{tab:daily-survey}). 
This left us with 804 valid responses. 
Three researchers performed thematic analysis on this data. 
Initially, each of the coders came up with their own set of codes. 
After discussion, we converged on a set of 52 codes. 
Through discussion, we reduced this subset of 52 codes to a list of 21 themes. 
Two of the researchers coded a sample (159 items, or 20\%) into the set of 52 codes. 
Coders had almost perfect agreement despite the large number of codes (Cohen's $\kappa = 0.82$)~\cite{viera2005understanding}. 
One of the coders coded the rest of the items. 
Due to the uneven number of codes associated with each user, as some did not regularly fill out the daily survey, we merged the 10-day codes for each person into a single set. 
Consequently, as we present our results, ``theme (10/41)'' means that out of 41 users, 10 users' responses contained at least one response belonging to the theme. 
Each user may have given such a response anywhere from 1 to \studydurationindays{} times over \studydurationindays{} days.

\subsubsection{Interview}
Similarly to the open-ended survey responses, we performed thematic analysis on the interview responses. 
One of the researchers performed an initial analysis and came up with 35 codes from the interviews. 
Then, this researcher discussed the codes and corresponding quotes with other researchers. 
After resolving disagreements, we were left with 24 codes from the interviews. As most of them were related to the daily survey themes, we merged the two sets. 

\section{Results}
Based on our research questions, we present results consistent across daily survey themes, interaction trace analysis, and interview themes. Note that the daily survey and interview quotes are presented in the format ``U1 (survey)'' and ``U1 (interview)''. 

\subsection{RQ1: Content Discovery}
\label{sec:content_discovery}
\begin{figure*}[t]
    \centering
    \includegraphics[width=0.8\textwidth]{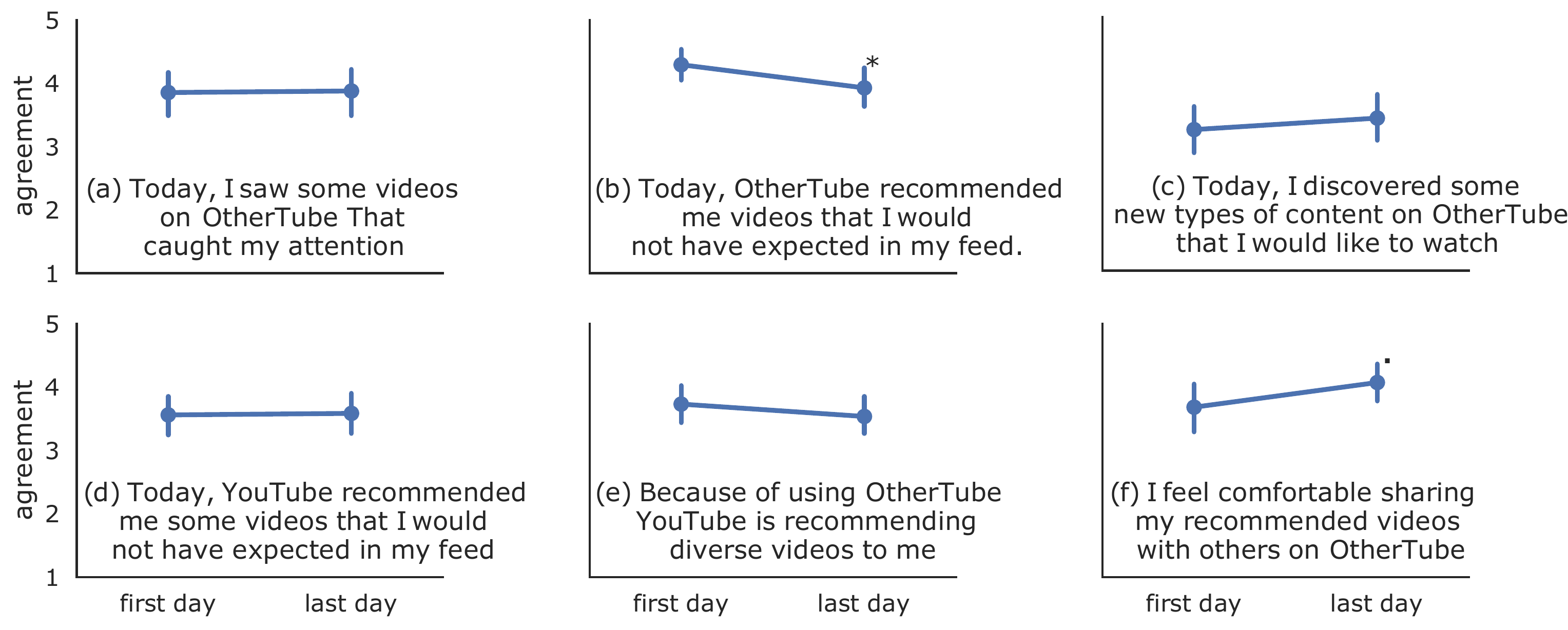}
    \caption{Mean with 95\% CI of participants' responses to the daily survey Likert items from the first and last days of the study. We also performed a Mann-Whitney U test comparing responses on the first and last days. In (b), * indicates $p<0.05$; in (f), \textbf{.} indicates $p<0.10$.}
    \label{fig:daily}
    \Description[Mean test results for the Likert items]{Only for item b and item f, the tests show significance.}
\end{figure*}

In the daily survey, the majority of participants agreed with the statement (mode: 4, ``somewhat agree'') saying that they saw some videos on \sys{} that caught their attention.
Further analysis shows that this perception did not change between the first and last days. 
This result is shown in Figure~\ref{fig:daily}-(a), where a Mann-Whitney U test shows no significant difference.
Our analysis revealed two themes that illustrate how users utilized \sys{} to discover content: (i) users developed new interests, and (ii) users rediscovered content they used to like. We present these themes below.

\subsubsection{\textbf{Users developed new interests}}
In the daily survey, a majority of our participants ($n=36/\studyparticipantcount{}$) mentioned finding new interests. 
Users found interests in multiple ways. Sometimes, they found new interests out of curiosity (\inlineqt{U1 (survey)}{I had no clue what it [a video that caught their attention] was about but just curious on the content}). 
Other times, new content led to the development of a new interest due to its usefulness (\inlineqt{U2 (survey)}{... A video about investing for your kids' future. I have a newborn and want to do that for her}).
There were also cases where users found new content that fit within their existing interests (\inlineqt{U12 (survey)}{Relaxing Video Game Music in a Cozy Room (Nintendo 64) [caught my attention]). I enjoy similar relaxing music playlists}.
Some users also watched content for the sake of exploration (\inlineqt{U17 (survey)}{I don't normally listen to that [classical music] and it was a welcome change}).
Interviewees also responded similarly, with one mentioning how they bumped into new interests on \sys{}.
Merely encountering the embedded interface of \sys{} could trigger changes in consumption behaviour.


\qt{U17 (interview)}{What I like about it is that some it forced me out of my comfort zone of what I would just do ... It makes you to stop and, like, look and think before you decide to make a different choice.}


We also noticed that participants were able to recognize profiles that had similar tastes to their own 
(\inlineqt{U20 (survey)}{I came across a user who had watched videos that were my interest. He was interested in computers and video games. So, I liked those videos too}).
This tendency demonstrates that one can find not only interesting videos using \sys{}, but also people one can connect with, due to similarities in the content that they watch. 
\sys's additional functions to retrieve more videos from a particular user can be useful to further explore a person's preferred content. 
Analyzing the number of clicks on More from This User and the number of clicks on videos users watched, we found a statistically significant positive correlation (Spearman's $\rho=0.42$, $p<0.001$). 
This pattern suggests that finding a profile based on similarities in taste could be a new way to find and watch new content---in essence, subscribing to other viewers, not creators, to follow their content consumption patterns.
We discuss these implications further in Section~\ref{dis:content}.

\subsubsection{\textbf{Users rediscovered content they used to like}}
Using \sys{}, some users ($n=11/41$) found content they used to like (\inlineqt{U11 (survey)}{old throwback videos from my childhood [caught my attention]}). They also found channels they used to like (\inlineqt{U3 (survey)}{a video game reviewer, Zero Punctuation, popped up in one of the \sys{} recommendations. So seeing that was pretty nostalgic.}).

Apart from these themes, our participant interviews further revealed users' issues with finding strangers with similar interests. 
For example, while users' About Me details were sometimes useful, they wanted more features, such as filtering users by demographic or interest (\inlineqt{U26 (interview)}{Their favorite content creator, maybe your favorite video.}).

\noindent\paragraph{\textbf{Summary:}} By browsing \sys{}, participants both found new interests and rediscovered old ones.
They found new interests out of curiosity, usefulness, or desire to explore.
Items caught participants' attention similarly on both the first and final days of the study.
Finally, participants also recognized others who had tastes similar to their own.
 
\begin{table}[t]
\centering
\footnotesize
\begin{tabular}{rll}
                                     \multicolumn{3}{r}{\textbf{\textit{Click Count per Day (Another Persona)}}}       \\
\hline
& \textbf{$\beta$}   & \textbf{std. Err.}           \\
\hline
(Intercept) & 5.5024*** & 1.0936 \\
Age & -0.0284* & 0.0135 \\
Gender[Male] & 0.2592 & 0.2303 \\
Race[Black or African American] & -1.2113 & 0.7136 \\
Race[Hispanic or Latino/a/x] & -0.1595 & 0.3895 \\
Race[White] & 0.2526 & 0.2412 \\
Daily Browsing Length[less than 1 hr] & -0.6087 & 0.3839 \\
Daily Browsing Length[less than 2 hrs] & -0.1536 & 0.3628 \\
Daily Browsing Length[greater than 2 hrs] & 0.1773 & 0.4109  \\
Need for Self-Reflection & -0.4149*** & 0.2108  \\
Insight & -0.0829 & 0.1730  \\
Day & -0.0453* & 0.0185 \\
\# of Users Available on the Day  & -0.0083 & 0.009 \\
\hline
Dispersion &  0.9119 & 0.1244\\
log.Lik       &  \multicolumn{2}{c}{-721.8501}\\
N = 280 & \multicolumn{2}{r}{* p\textless{}.05, ** p\textless{}.01, *** p\textless{}.001}\\
\hline
\end{tabular}
\caption{The mixed-effects negative binomial model for daily click count on Another Persona in Figure~\ref{fig:ot}. In this model, user is a random-effect variable. User demographics, their browsing habits, their need for self-reflection, the day when the clicks were counted, and the number of unique stranger data sets available to browse on \sys{} on the day when the clicks were counted are fixed-effect variables. The estimated negative binomial regression coefficient $\beta$ is the difference in the logarithms of expected counts of the response variable due to a one-unit change in the predictor variable.}\label{tab:reg}
\vspace{-20pt}
\end{table}

\subsection{RQ2: Factors Affecting Interaction and Engagement}
To investigate what factors affected users' interactions with \sys{}, we modeled users' click activity on Another Persona over time using a mixed-effects negative binomial model.
We considered several factors in this model, including users' demography, their need for self-reflection, their YouTube browsing length, and the number of unique profiles available to browse on a given day.
Table~\ref{tab:reg} shows the result.
We find that older users tend to interact less with \sys{} when browsing different personas ($\beta=-0.03, p<0.05$).
Our users with a greater need for self-reflection also interacted less with \sys{} ($\beta=-0.41, p<0.001$).
This result was particularly interesting because we hypothesized that those who believe that they need reflection would use \sys{} more than those who did not have such thoughts, given our goal of facilitating reflection behind the tool. 
Perhaps our design will be more effective and promising for those without an explicit desire for self-reflection. 
Additionally, as the days passed, users interacted less with the tool ($\beta=-0.05, p<0.05$).
This decrease could have been caused by seeing the same profiles repeatedly, due to the limited number of users in the study (\inlineqt{U4 (survey)}{[On day 4] i haven't seen many changes. [On day 6] no same as yesterday.}).
To understand users' engagement with the videos, we analyzed users' clicks on videos.
We found that most users ($n=37/41$) watched at least one video on \sys{}.
Overall, participants collectively clicked on 6\% or 398 videos out of all the collected videos on \sys{}.

While demographic attributes and other environmental factors can affect interaction, it might have been the content itself that encouraged or discouraged  interaction with content from \sys{}. 
Analysis of users' daily survey responses revealed content-related factors that discouraged users from watching videos; these are discussed below.

\subsubsection{\textbf{Users hesitated to watch content that was not of interest to them}}
A majority of our users ($n=25/41$) mentioned that they did not watch content because the content did not relate to their interests (\inlineqt{U40 (survey)}{I didn't find them [videos I hesitated to watch] interesting}). 
Some users referred to such content as ``boring'' (\inlineqt{U35 (survey)}{seemed boring or predictable}) or ``complicated'' (\inlineqt{U36 (survey)}{looked boring, or too techy and complicated}).
In the interview, one of the participants further elaborated on how their motivation for browsing YouTube at times affects whether they would engage with certain type of content (\inlineqt{U15 (interview)}{Sometimes I don't want to be educated. I just want to enjoy. Other people want maybe to learn about the history of Macedonia or something. I don't sometimes want to learn that.}).

\subsubsection{\textbf{Users hesitated to watch content that was not helpful}}
In contrast, content utility was the main factor for some users' ($n=6/41$) engagement with content in \sys{} (\inlineqt{U2 (survey)}{I just didn't click the ones that seemed pointless and not helpful to my personal development or learning}).
Some expressed hesitation in terms of wasted time (\inlineqt{U26 (survey)}{They could be a waste of time to watch if I don't like them}).

\subsubsection{\textbf{Users hesitated to watch disturbing and offensive content}}
Some users ($n=7/41$) were also hesitant to watch content that seemed disturbing or offensive to them.
Various types of content could fall into this category, including horror videos (\inlineqt{U13 (survey)}{I didn't want to watch horror}), sexually charged videos (\inlineqt{U38 (survey)}{I didn't want to watch videos that seemed strangely sexually charged, like mouth ASMRs}) and uncomfortable topics (\inlineqt{U39 (survey)}{I was disturbed by videos in uncomfortable topics I wouldn't want to think about, such as, physiological anomalies and unsolved murder mysteries}).

\subsubsection{\textbf{Users hesitated to watch clickbait-esque content}}
Clickbait was one reason some users supplied for not watching content ($n=4/41$).
Users used both video titles (\inlineqt{U16 (survey)}{Titles were terrible and looked like trash}) and thumbnails (\inlineqt{U38 (survey)}{I was hesitant to watch videos with objectifying thumbnail pictures of women}) to rule out this kind of content. 
 
Some of the barriers discussed so far may reduce the overall effectiveness of our approach, as users may consider the content displayed by \sys{} to consist of low-quality (e.g., spam, clickbait-esque videos) or even explicit (e.g., profanity) items. 
Future designs could introduce some mechanisms (e.g., automated spam filtering) to mitigate such perceptions.

\subsubsection{\textbf{Users hesitated to watch videos for fear of an interaction effect on their YouTube recommendations}}
A few of our participants ($n=4/41$) were hesitant to watch content because they assumed doing so would affect their future YouTube recommendations (\inlineqt{U25 (survey)}{I saw an interesting video on Netscape, but didn't click given it's from a new channel I'm not familiar with, but also being aware that the YT algorithm is going to try and keep me engaged by literally suggesting more of said video}). 
Some of our interviewees further revealed that they deal with content that they do not want to watch by blocking it. Similar blocking mechanisms could be added to \sys{} to filter out content that users do not want to see. We leave this for future exploration.

\qt{U39 (interview)}{I curated it [YouTube] fairly carefully. Occasionally I would get some random stuff [on YouTube recommendations]. Usually I just click on the little dots [a button on YouTube videos with a drop-down menu to block content users do not want to be recommended] and I'd say `don't recommend this channel'.}


This issue raises an interesting question of how users' engagement from watching content on \sys{} could affect their future recommendations from YouTube.
In light of this, should designers make \sys{} available within a sandbox or incognito viewing mode? 
Sandboxing may encourage some users concerned about impacting the recommendation algorithm to interact and explore more content.
 
\paragraph{\textbf{Summary:}} Our analyses show that factors such as lower age and lower need for self-reflection positively affected users' daily interaction with \sys{}.
However, participants' interaction also subsided as days passed.
We additionally found that 90\% of our participants watched at least one video on \sys{}.
Some content factors also discouraged engagement, including lack of relevance to users' interests, lack of utility, and offensive or clickbait-esque nature. 

\subsection{RQ3-1: Self-Reflection}
Using \sys{}, our participants were able to reflect on themselves from various perspectives. We found the following themes here: (i) understanding of own interests and their uniqueness, (ii) feelings of belonging from seeing users with similar interests, and (iii) feelings of superiority from comparing content. These patterns help us understand the kinds of self-reflection that the system can facilitate. We illustrate these themes below.

\subsubsection{\textbf{Users understood more about their own interests}}
One of the effects of using \sys{} seems to be users ($n=14/41$) understanding their own interests, what they like (\inlineqt{U41 (survey)}{I learn that I have interest in cooking after watching some cooking videos}), and what they do not like (\inlineqt{U15 (survey)}{I am not interested in animation or anything about computers or how computers work}). 
Some also realized their preferences in terms of video length (\inlineqt{U16 (survey)}{I like short and fun videos, nothing too long or too serious.}).
Others realized that their identity influences their interests (\inlineqt{U29 (survey)}{I saw my tastes to be very related to my identify [sic] and cultural background}).

After seeing strangers' YouTube recommendations, a majority of our participants ($n=29/41$) further realized that their interests were unique compared to those of others (\inlineqt{U9 (survey)}{I learned how unique and distinct my taste and preferences are}).
Some thought that their interests on YouTube were very narrow or selective compared to those of others.
Some of them defined this narrowness in terms of topics of interest (\inlineqt{U26 (survey)}{I mainly focus just on gaming while other people hit a lot more genres}).
These quotes show users gaining awareness about their limited interests when they can see beyond YouTube's filtered video suggestions.
 
\subsubsection{\textbf{Seeing others with similar interests gave users feelings of belonging }}\label{sec-self-ref3}
Users were surprised to find strangers with similar interests (\inlineqt{U29 (survey)}{I was surprised to see another person also to be having similar tastes as me, it was like seeing my own watch history}).
Seeing others with interests similar to their own, many of them ($n=22/41$) realized that they are not so different from others (\inlineqt{U5 (survey)}{I continue to learn that we aren't so different. I have always thought that from a gender perspective, but age is really where I feel my eyes are being opened.}).
Some found commonality among similar demographics (\inlineqt{U19 (survey)}{People near my age enjoy similar content}).
Others found commonality with different demographics (\inlineqt{U2 (survey)}{I like the same videos as a woman in her 50's which made me laugh because I'm 28}).
Some also expressed their alignment in interests with particular demographics over others (\inlineqt{U13 (survey)}{I learned that I am more likely to share interests with older women than men of my age}).
Some users expressed both their uniqueness and their differences as part of being normal (\inlineqt{U5 (survey)}{I think that I am fairly normal or that weird is normal. Everyone has their own tastes.}).
To see more from strangers with similar interests, \sys{} provided users with a pin button that allowed them to save other profiles for later perusal. 
We found that some of the participants ($n=8/41$) indeed pinned some strangers ($n=18/41$).

Our interviews further revealed that upon seeing users with similar interests, a few were interested in communicating with them, potentially to recommend them other interests (\inlineqt{U17 (interview)}{I think it'd be interesting to maybe even have that kind of connection where you're sharing. I don't want to say like you're friending someone on Facebook. But maybe sharing of content ... kind of being able to make a recommendation saying if you like this, maybe you like that.}).
However, at the same time, others were not interested in communicating for fear of toxic interactions. The quote below illustrates this issue.

\qt{U39 (interview)}{But I've been on YouTube for years ... my impression is that overall, YouTube is an extremely toxic place. And the only way I managed to avoid the toxicity is by blocking creators who I find to be like bad people and not reading the comments section. Some of the most innocuous videos will have thumbs down on them for like, I don't know what reason. So I don’t know if I want to connect with those people [who watch such videos]... And I thought that the way that you guys were doing it by making it impersonal was kind of good. Because then nobody could spew profanities at each other.}

\subsubsection{\textbf{Some users felt superior when comparing their content to others'}}
Although not many, a few of our participants ($n=6/41$) also reported feelings of superiority after seeing others' recommendations (\inlineqt{U16 (survey)}{Apparently, my taste is better than most people's.}).
Some of them attributed this sense of superiority to the lack of variety in others' recommendations
(\inlineqt{U19 (survey)}{I have more tastes than this lady. I seek out more content.}).
Manifestations of superiority go against our goal of understanding others and could be detrimental
(\inlineqt{U6 (survey)}{younger crowds listen to stupid stuff}).
 
\paragraph{\textbf{Summary:}} Using \sys{}, participants reflected on themselves and their interests.
When participants found others with interests matching their own, they experienced a feeling of belonging, and vice versa.
Comparing interests with others, some also felt superior to others.

\subsection{RQ3-2: Understanding of Others}
After seeing strangers' recommendations, participants expressed what they learned and how \sys{} increased their understanding of others. 
Some were surprised by the content others watch, while others were surprised by the fact that their preconceived notions about certain demographics did not always match. 
Users also discovered diversity in strangers' recommendations. We illustrate these themes below.

\subsubsection{\textbf{Users were surprised by some of the content others watch}}
When browsing their \sys{} feeds, some users ($n=11/41$) were surprised to see certain content. 
Sometimes, 
users were surprised by content when it came from a particular profile or demographic they did not expect it to come from 
(\inlineqt{U5 (survey)}{I found it very cool that someone in their 40s was still watching Olivia Rodrigo. Makes me feel better about aging haha}).
However, comparing users' responses over time, this surprise seems to have gradually subsided. 
Indeed, Figure~\ref{fig:daily}(b) shows a decrease in perceptions of seeing unexpected content on \sys{}, and this difference was statistically significant ($p<0.05$).
This result may indicate that participants got used to the recommended videos from \sys{}.
However, it is worth noting that it is unknown whether this trend would have arisen even with users being exposed to completely new sets of profiles every day; the limited number of profiles available could have contributed to this trend. 
Still, the mean of the responses in Figure~\ref{fig:daily}(b) did not fall below ``somewhat agree'' (4) on users' final days using \sys{}.
 
\subsubsection{\textbf{Users were surprised to see that certain stereotypes do not reflect reality}}
A few of our participants ($n=6/41$) were surprised to see that some stereotypes did not match reality (\inlineqt{U40 (survey)}{... It [different interests among people] challenges some stereotypes as I see things on their feed that I wouldn't have expected}).
Some found users of certain ages watching videos they would not expect them to watch (\inlineqt{U5 (survey)}{I continue to be surprised by music recommendations in particular. A lot of older persons getting younger artists and some younger persons recommended classic rock/pop}).
Some also were surprised in terms of gender stereotypes (\inlineqt{U8 (survey)}{I saw a male user watching some homey vlogs and that was a bit surprising}).

 
\subsubsection{\textbf{Users learned something particular about a person or a demographic, which could be stereotypical}}
Using \sys{}, about half of our participants ($n=17/41$) mentioned learning about others' lives by watching the recommendations that strangers received (\inlineqt{U29 (survey)}{I learnt how you can tell about what is happening in somebody's life by looking at the videos they are watching.}).
Some participants discovered what a particular demographic might be interested in (\inlineqt{U13 (survey)}{Men may be more into horror} and \inlineqt{U41 (survey)}{40s and older watch more health related videos}).
While some people experienced a feeling of superiority when learning about others' interests, others had the opposite reaction (\inlineqt{U29 (survey)}{I learned to respect other's choices}).

Meanwhile, some users ($n=5/41$) also found that stereotypes about the kinds of content particular demographics prefer matched what they saw (\inlineqt{U40 (survey)}{... It reaffirms some stereotypes that I had, as some videos are expected}).
Users most often mentioned stereotypes about age, particularly for younger people (\inlineqt{U5 (survey)}{Today was much more in line with expectations. A teenager recommended some Pokemon videos, 20s getting anime.}).
Some noticed similar recommendations related to ethnicity (\inlineqt{U38 (survey)}{I learned than ethnic Americans likely have video interests in their ethnic culture}).
Some also employed stereotypes to infer strangers' ages (\inlineqt{U5 (survey)}{Despite keeping their age hidden, my first profile today was clearly a kid/teen}).
One obstacle in understanding more about others could have been that some users kept their demographic information hidden, as allowed by \sys{} (\inlineqt{U41 (survey)}{several people didn't put their demographics. So I don't know their age group, gender.}).

\subsubsection{\textbf{Users discovered diversity in strangers' interests}}
Some of the participants ($n=10/41$) liked the diversity of the content they found on \sys{} (\inlineqt{U11 (survey)}{I like the diversity of participants and the content of the videos}).
They found diversity within particular strangers' feeds (\inlineqt{U3 (survey)}{... even within personas, it's getting harder to pin down a common thread. The recommendations can be pretty diverse.}).
Users also found diversity across demographics (\inlineqt{U4 (survey)}{Every age group is different and they all post different material on here.}).
 
Apart from reflecting on self and learning about others, nearly half of our participants ($n=18/41$) learned something about YouTube content broadly or what is trending on YouTube using \sys{}.
Some learned about common interests of people on YouTube (\inlineqt{U11 (survey)}{They [\sys{} users] all have fun videos}).
Others learned of new phenomena on YouTube (\inlineqt{U18 (survey)}{People are using YouTube as a news source with more regularity}).
Some discovered the popularity of certain content genres on YouTube (\inlineqt{U17 (survey)}{K-pop is massively popular. even more than I imagined}).
Some also learned about the popularity of content they already watch on YouTube (\inlineqt{U25 (survey)}{Realized some content I watch is actually popular.}), or about how the YouTube algorithm can affect their interests (\inlineqt{U13 (survey)}{When I don't have my own tastes in videos, Youtube shapes my tastes}).
 
\textbf{Summary:} After using \sys{} for \studydurationindays{} days, participants learned several things about others.
Some of these things surprised users, particularly when their preconceived ideas did not match what they learned.
Participants realized how diverse people's interests can be on YouTube.
 
 
\section{Discussion}
This study presents the utility of exchanging YouTube recommendations with strangers. 
RQ1 reveals unique ways in which this can facilitate content discovery.
RQ2 identifies demographics that might be interested in browsing recommendations catered to strangers and watching videos from them.
Finally, we saw how users reflected on themselves and others through RQ3.

\subsection{RQ1 \& RQ2: Content Discovery, Interaction, and Engagement}
\label{dis:content}
While YouTube provides some features for finding new content (e.g., search or subscription, exploring trending videos), diversification of YouTube recommendations is still challenging. It is especially hard to recommend diverse content to users based solely on their watch history. As one of our participants noted, 
\qt{U25 (interview)}{I sometimes want to find new songs ... it [YouTube] just kind of recycles a bunch of songs that I've heard ... In that instance, I actually have to basically open an incognito tab and pretend I'm just a random nobody to get actually novel or interesting music recommendations ... If I were to look for something new, I probably wouldn't be using my existing Google account essentially, rather try pretending to be a new person.}
In a way, our design emulates what this user has to do to get novel recommendations.
Use of \sys{} shows some promising results, with participants in the study both developing new interests and rediscovering prior interests after seeing content from strangers.
As our extension embeds diverse recommendations from the content available on YouTube, it could also act as a nudge for people to find new interests.
At the same time, our results also illuminate limits in the types of diverse content that can be shown.
For example, recommending offensive or excessively unfamiliar content could make people more hesitant to watch new kinds of content.
Consequently, there is room for improvement in designing systems for exchanging recommendations with strangers, especially in assisting users to find people with similar interests or showing profiles not completely randomly, but based on a few similarities, to prevent adverse reactions from users.
We could devise intelligent algorithms based on users' profiles to show content they may like. 
Additionally, we could computationally filter out spam content, like clickbait videos.
Others may need more case-by-case input (e.g., content on unfamiliar topics).
\additionminor{Based on our RQ1 results, future work could introduce two browsing modes---one for developing new interests and another for discovering old interests---to facilitate better content discovery. In each case, the system could provide suggestions based on their specific goal. 
For example, when browsing for old interests, users can specify time window that the system could use to search recommended videos from that period. Furthermore, users' lack of interaction could also be used to improve future suggestions. For example, when people choose not to interact with a stranger's profile, profiles that are similar to the profile could be assigned lower rankings, while dissimilar profiles could be given priority for later suggestions.}
 
\additionminor{
Our study was devised to test how users may react to a system for exchanging recommendations with strangers.
Therefore, users' reduced engagement over time is understandable in light of the small number of profiles available to browse, as seen in RQ2.
Nevertheless, it remains unclear how users would react with a larger pool of participants. 
New problems may arise in such an environment. 
For example, effectively browsing or navigating through a large number of strangers until they find content that interests them can be an overwhelming task for users, especially in the long run.
Perhaps new navigational techniques can be provided to the users to alleviate this issue. For example, if we allow users to filter strangers by various criteria (e.g., music interest and live video watching habit), it would be easier to browse within the group that fall under those criteria.
}
Overall, our work lays a foundation in this domain for exchanging recommendations between strangers, and further exploration is still needed.

\subsection{RQ3: Understanding of Oneself and Others}
As our results suggest, \sys{} helps users understand more about their interests and compare those interests with others' on YouTube.
This understanding could positively impact users if they experience a feeling of belonging by seeing others with similar interests.
Then again, the opposite can also happen; that is, if users are not able to find others with similar interests, they may instead feel disconnected, isolated, or atypical.
A potential solution to address the opposite effect is to design the system to suggest recommendations from users with some common ground (e.g., someone from the same demographic group, someone with similar taste) and monitor their interactions (e.g., video clicks).
However, pairings of extremely similar users should be avoided, as this may lead users back into their filter bubbles.
Understanding the right amount of similarity and dissimilarity from which users can easily relate to others while still learning something is an interesting challenge.
Being able to to search or filter users by demography (e.g., age range) and interests (e.g., favorite videos and content creators), as opposed to randomly suggesting profiles, may be another option for improving users' experiences with \sys{}.


Our results also suggest potential in creating social connections between strangers with common interests.
\additionminor{Our feature allowing users to pin strangers already act as a ``subscribe to another viewer'' type of function. One future improvement of this feature could be helping user realize how recommendation from their subscriptions change over time. To extend the subscribe feature to a two-way connection,}
future work would need to consider the case of opposing desires (users who want to connect vs. those who do not).
Perhaps designers can provide limited options for connecting pairs of strangers. 
For example, the system could unlock messaging options between two users only if they have multiple shared interests and want to connect~\cite{doris2013political}.
Alternatively, communication can be centered around content, such as by allowing users to recommend videos directly and respond to recommendations in a minimal fashion (e.g., using an emoji). 
\additionminor{In addition to making connections, one potential direction for research lies in designing interventions to nudge self-reflection. Users' activities, including browsing and watching patterns, can be used to trigger such interventions. For example, questions can be devised to help users reflect on changes in their browsing habits, reflect on connections with strangers, and encourage empathy-building with others. Or, notifications can be posted to users summarizing their usage information, such as, the number of strangers' profiles visited and the number of videos watched.}
Overall, future works could explore these directions for supporting reflection by exchanging recommendations with strangers.



 
\subsection{Design Implications}
On platforms such as YouTube, where recommendation algorithms contribute to shaping users' tastes and potentially trap users in filter bubbles for the sake of engagement, our approach of exchanging recommendations shows new ways to expand the scope of content discovery and improve reflection.
For content discovery, existing systems already adopt multiple approaches to provide users alternative recommendations to explore.
For example, apart from the site's personalized recommendations, \textit{Trending on YouTube} aims to promote exploration into ``videos that a wide range of viewers would find interesting''~\cite{Trending70online}.
Meanwhile, services like Spotify suggest content through features like \textit{Discover Weekly}, based on the interests of other users with preferences similar to one's own~\cite{HowSpoti80online}.
In contrast, our approach does not aggregate interests from any particular group; rather, it promotes exploration into other individual users' interests.
Because we use other individuals' YouTube recommendations, our approach is fundamentally unlike YouTube's trending recommendations.
As companies such as Alphabet, which owns YouTube, become keen to provide users with more opportunities to curate and expand their interests~\cite{KeenExpa82online}, our approach may provide such an opportunity.
\additionminor{In particular, this design creates opportunities for content consumers to follow their fellow consumers, instead of following only content creators.}
Aside from this, our approach also has the potential to fill a gap in the social functionality of systems like YouTube by creating connections between strangers based on similarity in interests.
\additionminor{Here, following strangers' recommendations can already act as a channel of communication, like a feedthrough mechanism~\cite{dix1994computer}.
On top of that, we may need to devise new approaches for communication.
Furthermore, by using following-follower connections as a weak form of social tie, recommender systems can learn more about users by examining who watches whose recommendations and use such information to improve recommendations.
Our design can also counteract filter bubbles in other settings, unlike solutions catering to only political echo chamber settings~\cite{munson2013encouraging}.
For example, merchants like Amazon can swap recommendations to promote new content to a user.
}


\subsection{Ethical Considerations}
We took several steps in our design and our study to minimize potential harm.
First, we collected only data that was needed for the tool and study. These included recommendations from users' YouTube homepages and their activity within the extension: the embedded content on the YouTube homepage, the extension's browser action page, and the extension's options page.
Second, we set the extension's content rating to ``mature'' on the Chrome Web Store to remind users that they could see explicit content in others' YouTube recommendations.
Third, we explicitly asked users to not share identifiable information in their profile fields to reduce the risk of an information leak (see Figure~\ref{fig:ot-profile}).
In case users were uncomfortable sharing any demographic details with others, the extension defaulted to not sharing any information; users manually chose what information to share with others. 
Similarly, we gave users the option to remove recommended videos from the collected set if they felt uncomfortable sharing them with others.
 
\subsection{Limitations}
Our study is not without limitations.
First, we had a limited number of participants from the U.S. with a limited demographic distribution.
Therefore, we cannot account for scenarios in which users live in different nations and speak different languages.
Additionally, while many users signed up for the study, only a select few eventually participated. Therefore, some self-selection bias exists.
These users could simply be those who are the most open to exchanging recommendations.
Furthermore, we recruited users who use the Chrome browser on a desktop or laptop computer (required for the \sys{} extension to work) and who use YouTube regularly.
These criteria likely excluded users who do not use YouTube regularly or who use it on a different platform (e.g., mobile phones).
Second, our implementation of \sys{} also introduced some limits to the study. For example, some participants hesitated to watch content because preview clips were unavailable.
One interview participant mentioned that they had trouble going back to a profile they had forgotten to pin.
This issue could have impacted their overall interaction and engagement with \sys.
Third, the first few participants from the first batch saw fewer profiles compared to other users; for comparison, the last person in the study had a maximum of 40 profiles to browse.
Therefore, users' responses to the daily survey were affected by the limited sample.
\additionminor{In addition to these limitations, in the future, studies can be devised to investigate the measurable impact of systems like \sys{} on users' consumption. For example, a within-subjects study could help us quantify the difference between users' interactions with \sys{} and YouTube recommendations.}
Future deployment with large samples could potentially resolve these issues.
 
 
\section{Conclusion}
In this work, we investigated the exchange of social recommendations with strangers as a tool to promote content discovery and reflection on social media sites like YouTube. 
Our investigation revealed factors that affect users' interaction and engagement with such a system.
Our work has implications for future exploration into exchanging personalized recommendations with strangers.
 

\begin{acks}
\additionminor{This paper would not be possible without our study participants. We also appreciate the valuable feedback we received from the anonymous reviewers, the members of the EchoLab at Virginia Tech, and the members of the Social Computing Lab at the University of Washington.}
\end{acks}

\bibliographystyle{ACM-Reference-Format}
\bibliography{sample-base}


\begin{thebibliography}{94}


\ifx \showCODEN    \undefined \def \showCODEN     #1{\unskip}     \fi
\ifx \showDOI      \undefined \def \showDOI       #1{#1}\fi
\ifx \showISBNx    \undefined \def \showISBNx     #1{\unskip}     \fi
\ifx \showISBNxiii \undefined \def \showISBNxiii  #1{\unskip}     \fi
\ifx \showISSN     \undefined \def \showISSN      #1{\unskip}     \fi
\ifx \showLCCN     \undefined \def \showLCCN      #1{\unskip}     \fi
\ifx \shownote     \undefined \def \shownote      #1{#1}          \fi
\ifx \showarticletitle \undefined \def \showarticletitle #1{#1}   \fi
\ifx \showURL      \undefined \def \showURL       {\relax}        \fi
\providecommand\bibfield[2]{#2}
\providecommand\bibinfo[2]{#2}
\providecommand\natexlab[1]{#1}
\providecommand\showeprint[2][]{arXiv:#2}

\bibitem[\protect\citeauthoryear{Adomavicius and Kwon}{Adomavicius and
  Kwon}{2009}]%
        {adomavicius2009toward}
\bibfield{author}{\bibinfo{person}{Gediminas Adomavicius} {and}
  \bibinfo{person}{YoungOk Kwon}.} \bibinfo{year}{2009}\natexlab{}.
\newblock \showarticletitle{Toward more diverse recommendations: Item
  re-ranking methods for recommender systems}. In
  \bibinfo{booktitle}{\emph{Workshop on Information Technologies and Systems}}.
  Citeseer.
\newblock


\bibitem[\protect\citeauthoryear{Ali, Foreman, Capasso, Jones, Tozan, and
  DiClemente}{Ali et~al\mbox{.}}{2020}]%
        {ali2020social}
\bibfield{author}{\bibinfo{person}{Shahmir~H Ali}, \bibinfo{person}{Joshua
  Foreman}, \bibinfo{person}{Ariadna Capasso}, \bibinfo{person}{Abbey~M Jones},
  \bibinfo{person}{Yesim Tozan}, {and} \bibinfo{person}{Ralph~J DiClemente}.}
  \bibinfo{year}{2020}\natexlab{}.
\newblock \showarticletitle{Social media as a recruitment platform for a
  nationwide online survey of COVID-19 knowledge, beliefs, and practices in the
  United States: methodology and feasibility analysis}.
\newblock \bibinfo{journal}{\emph{BMC medical research methodology}}
  \bibinfo{volume}{20} (\bibinfo{year}{2020}), \bibinfo{pages}{1--11}.
\newblock


\bibitem[\protect\citeauthoryear{Andr\'{e}, Schraefel, Dix, and
  White}{Andr\'{e} et~al\mbox{.}}{2011}]%
        {andre2011expressing}
\bibfield{author}{\bibinfo{person}{Paul Andr\'{e}}, \bibinfo{person}{M.~C.
  Schraefel}, \bibinfo{person}{Alan Dix}, {and} \bibinfo{person}{Ryen~W.
  White}.} \bibinfo{year}{2011}\natexlab{}.
\newblock \showarticletitle{Expressing Well-Being Online: Towards
  Self-Reflection and Social Awareness}. In
  \bibinfo{booktitle}{\emph{Proceedings of the 2011 IConference}} (Seattle,
  Washington, USA) \emph{(\bibinfo{series}{iConference '11})}.
  \bibinfo{publisher}{Association for Computing Machinery},
  \bibinfo{address}{New York, NY, USA}, \bibinfo{pages}{114–121}.
\newblock
\showISBNx{9781450301213}
\urldef\tempurl%
\url{https://doi.org/10.1145/1940761.1940777}
\showDOI{\tempurl}


\bibitem[\protect\citeauthoryear{Back, Stopfer, Vazire, Gaddis, Schmukle,
  Egloff, and Gosling}{Back et~al\mbox{.}}{2010}]%
        {back2010facebook}
\bibfield{author}{\bibinfo{person}{Mitja~D Back}, \bibinfo{person}{Juliane~M
  Stopfer}, \bibinfo{person}{Simine Vazire}, \bibinfo{person}{Sam Gaddis},
  \bibinfo{person}{Stefan~C Schmukle}, \bibinfo{person}{Boris Egloff}, {and}
  \bibinfo{person}{Samuel~D Gosling}.} \bibinfo{year}{2010}\natexlab{}.
\newblock \showarticletitle{Facebook profiles reflect actual personality, not
  self-idealization}.
\newblock \bibinfo{journal}{\emph{Psychological science}} \bibinfo{volume}{21},
  \bibinfo{number}{3} (\bibinfo{year}{2010}), \bibinfo{pages}{372--374}.
\newblock


\bibitem[\protect\citeauthoryear{Bae, Lim, Bang, and Kim}{Bae
  et~al\mbox{.}}{2014}]%
        {bae2014ripening}
\bibfield{author}{\bibinfo{person}{Jae-eul Bae}, \bibinfo{person}{Youn-kyung
  Lim}, \bibinfo{person}{Jin-bae Bang}, {and} \bibinfo{person}{Myung-suk Kim}.}
  \bibinfo{year}{2014}\natexlab{}.
\newblock \showarticletitle{Ripening Room: Designing Social Media for
  Self-Reflection in Self-Expression}. In \bibinfo{booktitle}{\emph{Proceedings
  of the 2014 Conference on Designing Interactive Systems}} (Vancouver, BC,
  Canada) \emph{(\bibinfo{series}{DIS '14})}. \bibinfo{publisher}{Association
  for Computing Machinery}, \bibinfo{address}{New York, NY, USA},
  \bibinfo{pages}{103–112}.
\newblock
\showISBNx{9781450329026}
\urldef\tempurl%
\url{https://doi.org/10.1145/2598510.2598567}
\showDOI{\tempurl}


\bibitem[\protect\citeauthoryear{Bales and Griswold}{Bales and
  Griswold}{2011}]%
        {bales2011interpersonal}
\bibfield{author}{\bibinfo{person}{Elizabeth Bales} {and}
  \bibinfo{person}{William Griswold}.} \bibinfo{year}{2011}\natexlab{}.
\newblock \showarticletitle{Interpersonal Informatics: Making Social Influence
  Visible}. In \bibinfo{booktitle}{\emph{CHI '11 Extended Abstracts on Human
  Factors in Computing Systems}} (Vancouver, BC, Canada)
  \emph{(\bibinfo{series}{CHI EA '11})}. \bibinfo{publisher}{Association for
  Computing Machinery}, \bibinfo{address}{New York, NY, USA},
  \bibinfo{pages}{2227–2232}.
\newblock
\showISBNx{9781450302685}
\urldef\tempurl%
\url{https://doi.org/10.1145/1979742.1979924}
\showDOI{\tempurl}


\bibitem[\protect\citeauthoryear{Baumer}{Baumer}{2015}]%
        {baumer2015reflective}
\bibfield{author}{\bibinfo{person}{Eric P~S Baumer}.}
  \bibinfo{year}{2015}\natexlab{}.
\newblock \showarticletitle{{Reflective informatics: conceptual dimensions for
  designing technologies of reflection}}. In
  \bibinfo{booktitle}{\emph{Proceedings of the 33rd Annual ACM Conference on
  Human Factors in Computing Systems}}. ACM, \bibinfo{publisher}{Association
  for Computing Machinery}, \bibinfo{address}{New York, NY, USA},
  \bibinfo{pages}{585--594}.
\newblock


\bibitem[\protect\citeauthoryear{Bessi, Zollo, Del~Vicario, Puliga, Scala,
  Caldarelli, Uzzi, and Quattrociocchi}{Bessi et~al\mbox{.}}{2016}]%
        {bessi2016users}
\bibfield{author}{\bibinfo{person}{Alessandro Bessi}, \bibinfo{person}{Fabiana
  Zollo}, \bibinfo{person}{Michela Del~Vicario}, \bibinfo{person}{Michelangelo
  Puliga}, \bibinfo{person}{Antonio Scala}, \bibinfo{person}{Guido Caldarelli},
  \bibinfo{person}{Brian Uzzi}, {and} \bibinfo{person}{Walter Quattrociocchi}.}
  \bibinfo{year}{2016}\natexlab{}.
\newblock \showarticletitle{Users polarization on Facebook and Youtube}.
\newblock \bibinfo{journal}{\emph{PloS one}} \bibinfo{volume}{11},
  \bibinfo{number}{8} (\bibinfo{year}{2016}), \bibinfo{pages}{e0159641}.
\newblock


\bibitem[\protect\citeauthoryear{Bhuiyan, Horning, Lee, and Mitra}{Bhuiyan
  et~al\mbox{.}}{2021}]%
        {bhuiyan2021nudgecred}
\bibfield{author}{\bibinfo{person}{Md~Momen Bhuiyan}, \bibinfo{person}{Michael
  Horning}, \bibinfo{person}{Sang~Won Lee}, {and} \bibinfo{person}{Tanushree
  Mitra}.} \bibinfo{year}{2021}\natexlab{}.
\newblock \showarticletitle{NudgeCred: Supporting News Credibility Assessment
  on Social Media Through Nudges}.
\newblock \bibinfo{journal}{\emph{Proceedings of the ACM on Human-Computer
  Interaction}} \bibinfo{volume}{5}, \bibinfo{number}{CSCW2}
  (\bibinfo{year}{2021}), \bibinfo{pages}{1--30}.
\newblock


\bibitem[\protect\citeauthoryear{Bhuiyan, Zhang, Vick, Horning, and
  Mitra}{Bhuiyan et~al\mbox{.}}{2018}]%
        {bhuiyan2018feedreflect}
\bibfield{author}{\bibinfo{person}{Md~Momen Bhuiyan}, \bibinfo{person}{Kexin
  Zhang}, \bibinfo{person}{Kelsey Vick}, \bibinfo{person}{Michael~A Horning},
  {and} \bibinfo{person}{Tanushree Mitra}.} \bibinfo{year}{2018}\natexlab{}.
\newblock \showarticletitle{FeedReflect: A Tool for Nudging Users to Assess
  News Credibility on Twitter}. In \bibinfo{booktitle}{\emph{Companion of the
  2018 ACM Conference on Computer Supported Cooperative Work and Social
  Computing}}. \bibinfo{pages}{205--208}.
\newblock


\bibitem[\protect\citeauthoryear{Boas, Christenson, and Glick}{Boas
  et~al\mbox{.}}{2020}]%
        {boas2020recruiting}
\bibfield{author}{\bibinfo{person}{Taylor~C Boas}, \bibinfo{person}{Dino~P
  Christenson}, {and} \bibinfo{person}{David~M Glick}.}
  \bibinfo{year}{2020}\natexlab{}.
\newblock \showarticletitle{Recruiting large online samples in the United
  States and India: Facebook, mechanical turk, and qualtrics}.
\newblock \bibinfo{journal}{\emph{Political Science Research and Methods}}
  \bibinfo{volume}{8}, \bibinfo{number}{2} (\bibinfo{year}{2020}),
  \bibinfo{pages}{232--250}.
\newblock


\bibitem[\protect\citeauthoryear{Bryant}{Bryant}{2020}]%
        {bryant2020youtube}
\bibfield{author}{\bibinfo{person}{Lauren~Valentino Bryant}.}
  \bibinfo{year}{2020}\natexlab{}.
\newblock \showarticletitle{The YouTube algorithm and the Alt-Right filter
  bubble}.
\newblock \bibinfo{journal}{\emph{Open Information Science}}
  \bibinfo{volume}{4}, \bibinfo{number}{1} (\bibinfo{year}{2020}),
  \bibinfo{pages}{85--90}.
\newblock


\bibitem[\protect\citeauthoryear{Burbach, Halbach, Ziefle, and
  {Calero{\^{A}} Valdez}}{Burbach et~al\mbox{.}}{2019}]%
        {Burbach2019}
\bibfield{author}{\bibinfo{person}{Laura Burbach}, \bibinfo{person}{Patrick
  Halbach}, \bibinfo{person}{Martina Ziefle}, {and}
  \bibinfo{person}{Andr{\'{e}} {Calero{\^{A}} Valdez}}.}
  \bibinfo{year}{2019}\natexlab{}.
\newblock \showarticletitle{{Bubble Trouble: Strategies Against Filter Bubbles
  in Online Social Networks}}.
\newblock \bibinfo{journal}{\emph{Lecture Notes in Computer Science (including
  subseries Lecture Notes in Artificial Intelligence and Lecture Notes in
  Bioinformatics)}}  \bibinfo{volume}{11582 LNCS} (\bibinfo{year}{2019}),
  \bibinfo{pages}{441--456}.
\newblock
\showISBNx{9783030222185}
\showISSN{16113349}
\urldef\tempurl%
\url{https://doi.org/10.1007/978-3-030-22219-2_33}
\showDOI{\tempurl}


\bibitem[\protect\citeauthoryear{Castells, Hurley, and Vargas}{Castells
  et~al\mbox{.}}{2015}]%
        {castells2015novelty}
\bibfield{author}{\bibinfo{person}{Pablo Castells}, \bibinfo{person}{Neil~J.
  Hurley}, {and} \bibinfo{person}{Saul Vargas}.}
  \bibinfo{year}{2015}\natexlab{}.
\newblock \bibinfo{booktitle}{\emph{Novelty and Diversity in Recommender
  Systems}}.
\newblock \bibinfo{publisher}{Springer US}, \bibinfo{address}{Boston, MA},
  \bibinfo{pages}{881--918}.
\newblock
\showISBNx{978-1-4899-7637-6}
\urldef\tempurl%
\url{https://doi.org/10.1007/978-1-4899-7637-6_26}
\showDOI{\tempurl}


\bibitem[\protect\citeauthoryear{Chen, Zhou, Zhu, and Xu}{Chen
  et~al\mbox{.}}{2012}]%
        {chen2012detecting}
\bibfield{author}{\bibinfo{person}{Ying Chen}, \bibinfo{person}{Yilu Zhou},
  \bibinfo{person}{Sencun Zhu}, {and} \bibinfo{person}{Heng Xu}.}
  \bibinfo{year}{2012}\natexlab{}.
\newblock \showarticletitle{Detecting Offensive Language in Social Media to
  Protect Adolescent Online Safety}. In \bibinfo{booktitle}{\emph{2012
  International Conference on Privacy, Security, Risk and Trust and 2012
  International Conference on Social Computing}}. IEEE,
  \bibinfo{pages}{71--80}.
\newblock
\urldef\tempurl%
\url{https://doi.org/10.1109/SocialCom-PASSAT.2012.55}
\showDOI{\tempurl}


\bibitem[\protect\citeauthoryear{Cohen and Syme}{Cohen and Syme}{1985}]%
        {cohen1985social}
\bibfield{author}{\bibinfo{person}{Sheldon~Ed Cohen} {and} \bibinfo{person}{SI
  Syme}.} \bibinfo{year}{1985}\natexlab{}.
\newblock \bibinfo{booktitle}{\emph{Social support and health.}}
\newblock \bibinfo{publisher}{Academic Press}.
\newblock


\bibitem[\protect\citeauthoryear{Consolvo, Everitt, Smith, and Landay}{Consolvo
  et~al\mbox{.}}{2006}]%
        {consolvo2006design}
\bibfield{author}{\bibinfo{person}{Sunny Consolvo}, \bibinfo{person}{Katherine
  Everitt}, \bibinfo{person}{Ian Smith}, {and} \bibinfo{person}{James~A
  Landay}.} \bibinfo{year}{2006}\natexlab{}.
\newblock \showarticletitle{Design requirements for technologies that encourage
  physical activity}. In \bibinfo{booktitle}{\emph{Proceedings of the SIGCHI
  conference on Human Factors in computing systems}}.
  \bibinfo{publisher}{Association for Computing Machinery},
  \bibinfo{address}{New York, NY, USA}, \bibinfo{pages}{457--466}.
\newblock


\bibitem[\protect\citeauthoryear{Consolvo, McDonald, and Landay}{Consolvo
  et~al\mbox{.}}{2009}]%
        {consolvo2009theory}
\bibfield{author}{\bibinfo{person}{Sunny Consolvo}, \bibinfo{person}{David~W.
  McDonald}, {and} \bibinfo{person}{James~A. Landay}.}
  \bibinfo{year}{2009}\natexlab{}.
\newblock \bibinfo{booktitle}{\emph{Theory-Driven Design Strategies for
  Technologies That Support Behavior Change in Everyday Life}}.
\newblock \bibinfo{publisher}{Association for Computing Machinery},
  \bibinfo{address}{New York, NY, USA}, \bibinfo{pages}{405–414}.
\newblock
\showISBNx{9781605582467}
\urldef\tempurl%
\url{https://doi.org/10.1145/1518701.1518766}
\showURL{%
\tempurl}


\bibitem[\protect\citeauthoryear{Covington, Adams, and Sargin}{Covington
  et~al\mbox{.}}{2016}]%
        {covington2016deep}
\bibfield{author}{\bibinfo{person}{Paul Covington}, \bibinfo{person}{Jay
  Adams}, {and} \bibinfo{person}{Emre Sargin}.}
  \bibinfo{year}{2016}\natexlab{}.
\newblock \showarticletitle{Deep Neural Networks for YouTube Recommendations}.
  In \bibinfo{booktitle}{\emph{Proceedings of the 10th ACM Conference on
  Recommender Systems}} (Boston, Massachusetts, USA)
  \emph{(\bibinfo{series}{RecSys '16})}. \bibinfo{publisher}{Association for
  Computing Machinery}, \bibinfo{address}{New York, NY, USA},
  \bibinfo{pages}{191–198}.
\newblock
\showISBNx{9781450340359}
\urldef\tempurl%
\url{https://doi.org/10.1145/2959100.2959190}
\showDOI{\tempurl}


\bibitem[\protect\citeauthoryear{Davidson, Liebald, Liu, Nandy, and {Van
  Vleet}}{Davidson et~al\mbox{.}}{2010}]%
        {Davidson2010}
\bibfield{author}{\bibinfo{person}{James Davidson}, \bibinfo{person}{Benjamin
  Liebald}, \bibinfo{person}{Junning Liu}, \bibinfo{person}{Palash Nandy},
  {and} \bibinfo{person}{Taylor {Van Vleet}}.} \bibinfo{year}{2010}\natexlab{}.
\newblock \showarticletitle{{The YouTube video recommendation system}}.
\newblock \bibinfo{journal}{\emph{RecSys'10 - Proceedings of the 4th ACM
  Conference on Recommender Systems}} (\bibinfo{year}{2010}),
  \bibinfo{pages}{293--296}.
\newblock
\showISBNx{9781450304429}
\urldef\tempurl%
\url{https://doi.org/10.1145/1864708.1864770}
\showDOI{\tempurl}


\bibitem[\protect\citeauthoryear{De~Choudhury, Gamon, Hoff, and
  Roseway}{De~Choudhury et~al\mbox{.}}{2013}]%
        {de2013moon}
\bibfield{author}{\bibinfo{person}{Munmun De~Choudhury},
  \bibinfo{person}{Michael Gamon}, \bibinfo{person}{Aaron Hoff}, {and}
  \bibinfo{person}{Asta Roseway}.} \bibinfo{year}{2013}\natexlab{}.
\newblock \showarticletitle{“Moon Phrases”: A social media faciliated tool
  for emotional reflection and wellness}. In \bibinfo{booktitle}{\emph{2013 7th
  International Conference on Pervasive Computing Technologies for Healthcare
  and Workshops}}. IEEE, \bibinfo{publisher}{European Alliance for Innovation},
  \bibinfo{pages}{41--44}.
\newblock


\bibitem[\protect\citeauthoryear{Dix}{Dix}{1994}]%
        {dix1994computer}
\bibfield{author}{\bibinfo{person}{Alan Dix}.} \bibinfo{year}{1994}\natexlab{}.
\newblock \showarticletitle{Computer supported cooperative work: a framework}.
\newblock In \bibinfo{booktitle}{\emph{Design issues in CSCW}}.
  \bibinfo{publisher}{Springer}, \bibinfo{pages}{9--26}.
\newblock


\bibitem[\protect\citeauthoryear{Doris-Down, Versee, and Gilbert}{Doris-Down
  et~al\mbox{.}}{2013}]%
        {doris2013political}
\bibfield{author}{\bibinfo{person}{Abraham Doris-Down}, \bibinfo{person}{Husayn
  Versee}, {and} \bibinfo{person}{Eric Gilbert}.}
  \bibinfo{year}{2013}\natexlab{}.
\newblock \showarticletitle{Political blend: an application designed to bring
  people together based on political differences}. In
  \bibinfo{booktitle}{\emph{Proceedings of the 6th International Conference on
  Communities and Technologies}}. \bibinfo{publisher}{Association for Computing
  Machinery}, \bibinfo{address}{New York, NY, USA}, \bibinfo{pages}{120--130}.
\newblock


\bibitem[\protect\citeauthoryear{Epstein, Ping, Fogarty, and Munson}{Epstein
  et~al\mbox{.}}{2015}]%
        {epstein2015lived}
\bibfield{author}{\bibinfo{person}{Daniel~A Epstein}, \bibinfo{person}{An
  Ping}, \bibinfo{person}{James Fogarty}, {and} \bibinfo{person}{Sean~A
  Munson}.} \bibinfo{year}{2015}\natexlab{}.
\newblock \showarticletitle{A lived informatics model of personal informatics}.
  In \bibinfo{booktitle}{\emph{Proceedings of the 2015 ACM International Joint
  Conference on Pervasive and Ubiquitous Computing}}.
  \bibinfo{publisher}{Association for Computing Machinery},
  \bibinfo{address}{New York, NY, USA}, \bibinfo{pages}{731--742}.
\newblock


\bibitem[\protect\citeauthoryear{Eslami, Aleyasen, Karahalios, Hamilton, and
  Sandvig}{Eslami et~al\mbox{.}}{2015a}]%
        {eslami2015feedvis}
\bibfield{author}{\bibinfo{person}{Motahhare Eslami},
  \bibinfo{person}{Amirhossein Aleyasen}, \bibinfo{person}{Karrie Karahalios},
  \bibinfo{person}{Kevin Hamilton}, {and} \bibinfo{person}{Christian Sandvig}.}
  \bibinfo{year}{2015}\natexlab{a}.
\newblock \showarticletitle{Feedvis: A path for exploring news feed curation
  algorithms}. In \bibinfo{booktitle}{\emph{Proceedings of the 18th acm
  conference companion on computer supported cooperative work \& social
  computing}}. \bibinfo{publisher}{Association for Computing Machinery},
  \bibinfo{address}{New York, NY, USA}, \bibinfo{pages}{65--68}.
\newblock


\bibitem[\protect\citeauthoryear{Eslami, Rickman, Vaccaro, Aleyasen, Vuong,
  Karahalios, Hamilton, and Sandvig}{Eslami et~al\mbox{.}}{2015b}]%
        {eslami2015always}
\bibfield{author}{\bibinfo{person}{Motahhare Eslami}, \bibinfo{person}{Aimee
  Rickman}, \bibinfo{person}{Kristen Vaccaro}, \bibinfo{person}{Amirhossein
  Aleyasen}, \bibinfo{person}{Andy Vuong}, \bibinfo{person}{Karrie Karahalios},
  \bibinfo{person}{Kevin Hamilton}, {and} \bibinfo{person}{Christian Sandvig}.}
  \bibinfo{year}{2015}\natexlab{b}.
\newblock \showarticletitle{" I always assumed that I wasn't really that close
  to [her]" Reasoning about Invisible Algorithms in News Feeds}. In
  \bibinfo{booktitle}{\emph{Proceedings of the 33rd annual ACM conference on
  human factors in computing systems}}. \bibinfo{publisher}{Association for
  Computing Machinery}, \bibinfo{address}{New York, NY, USA},
  \bibinfo{pages}{153--162}.
\newblock


\bibitem[\protect\citeauthoryear{Fan and Poole}{Fan and Poole}{2006}]%
        {fan2006personalization}
\bibfield{author}{\bibinfo{person}{Haiyan Fan} {and}
  \bibinfo{person}{Marshall~Scott Poole}.} \bibinfo{year}{2006}\natexlab{}.
\newblock \showarticletitle{What is personalization? Perspectives on the design
  and implementation of personalization in information systems}.
\newblock \bibinfo{journal}{\emph{Journal of Organizational Computing and
  Electronic Commerce}} \bibinfo{volume}{16}, \bibinfo{number}{3-4}
  (\bibinfo{year}{2006}), \bibinfo{pages}{179--202}.
\newblock


\bibitem[\protect\citeauthoryear{Festinger}{Festinger}{1954}]%
        {festinger1954theory}
\bibfield{author}{\bibinfo{person}{Leon Festinger}.}
  \bibinfo{year}{1954}\natexlab{}.
\newblock \showarticletitle{A theory of social comparison processes}.
\newblock \bibinfo{journal}{\emph{Human relations}} \bibinfo{volume}{7},
  \bibinfo{number}{2} (\bibinfo{year}{1954}), \bibinfo{pages}{117--140}.
\newblock


\bibitem[\protect\citeauthoryear{Feustel, Aggarwal, Lee, and Wilcox}{Feustel
  et~al\mbox{.}}{2018}]%
        {feustel2018people}
\bibfield{author}{\bibinfo{person}{Clayton Feustel}, \bibinfo{person}{Shyamak
  Aggarwal}, \bibinfo{person}{Bongshin Lee}, {and} \bibinfo{person}{Lauren
  Wilcox}.} \bibinfo{year}{2018}\natexlab{}.
\newblock \showarticletitle{People like me: Designing for reflection on
  aggregate cohort data in personal informatics systems}.
\newblock \bibinfo{journal}{\emph{Proceedings of the ACM on Interactive,
  Mobile, Wearable and Ubiquitous Technologies}} \bibinfo{volume}{2},
  \bibinfo{number}{3} (\bibinfo{year}{2018}), \bibinfo{pages}{1--21}.
\newblock


\bibitem[\protect\citeauthoryear{Gao}{Gao}{2012}]%
        {gao2012design}
\bibfield{author}{\bibinfo{person}{Feng Gao}.} \bibinfo{year}{2012}\natexlab{}.
\newblock \showarticletitle{Design for reflection on health behavior change}.
  In \bibinfo{booktitle}{\emph{Proceedings of the 2012 ACM international
  conference on Intelligent User Interfaces}}. \bibinfo{publisher}{Association
  for Computing Machinery}, \bibinfo{address}{New York, NY, USA},
  \bibinfo{pages}{379--382}.
\newblock


\bibitem[\protect\citeauthoryear{Garbett, Chatting, Wilkinson, Lee, and
  Kharrufa}{Garbett et~al\mbox{.}}{2018}]%
        {garbett2018thinkactive}
\bibfield{author}{\bibinfo{person}{Andrew Garbett}, \bibinfo{person}{David
  Chatting}, \bibinfo{person}{Gerard Wilkinson}, \bibinfo{person}{Clement Lee},
  {and} \bibinfo{person}{Ahmed Kharrufa}.} \bibinfo{year}{2018}\natexlab{}.
\newblock \bibinfo{booktitle}{\emph{ThinkActive: Designing for Pseudonymous
  Activity Tracking in the Classroom}}.
\newblock \bibinfo{publisher}{Association for Computing Machinery},
  \bibinfo{address}{New York, NY, USA}, \bibinfo{pages}{1–13}.
\newblock
\showISBNx{9781450356206}
\urldef\tempurl%
\url{https://doi.org/10.1145/3173574.3173581}
\showURL{%
\tempurl}


\bibitem[\protect\citeauthoryear{Gentile, Twenge, Freeman, and
  Campbell}{Gentile et~al\mbox{.}}{2012}]%
        {gentile2012effect}
\bibfield{author}{\bibinfo{person}{Brittany Gentile}, \bibinfo{person}{Jean~M
  Twenge}, \bibinfo{person}{Elise~C Freeman}, {and} \bibinfo{person}{W~Keith
  Campbell}.} \bibinfo{year}{2012}\natexlab{}.
\newblock \showarticletitle{The effect of social networking websites on
  positive self-views: An experimental investigation}.
\newblock \bibinfo{journal}{\emph{Computers in human behavior}}
  \bibinfo{volume}{28}, \bibinfo{number}{5} (\bibinfo{year}{2012}),
  \bibinfo{pages}{1929--1933}.
\newblock


\bibitem[\protect\citeauthoryear{Gershgorn}{Gershgorn}{2019}]%
        {HowSpoti80online}
\bibfield{author}{\bibinfo{person}{Dave Gershgorn}.}
  \bibinfo{year}{2019}\natexlab{}.
\newblock \bibinfo{title}{How Spotify’s Algorithm Knows Exactly What You Want
  to Listen To | by Dave Gershgorn | OneZero}.
\newblock
  \bibinfo{howpublished}{\url{https://onezero.medium.com/how-spotifys-algorithm-knows-exactly-what-you-want-to-listen-to-4b6991462c5c}}.
\newblock
\newblock
\shownote{(Accessed on 09/09/2021).}


\bibitem[\protect\citeauthoryear{Gillani, Yuan, Saveski, Vosoughi, and
  Roy}{Gillani et~al\mbox{.}}{2018}]%
        {Gillani2018}
\bibfield{author}{\bibinfo{person}{Nabeel Gillani}, \bibinfo{person}{Ann Yuan},
  \bibinfo{person}{Martin Saveski}, \bibinfo{person}{Soroush Vosoughi}, {and}
  \bibinfo{person}{Deb Roy}.} \bibinfo{year}{2018}\natexlab{}.
\newblock \showarticletitle{{Me, my echo chamber, and i: Introspection on
  social media polarization}}.
\newblock \bibinfo{journal}{\emph{The Web Conference 2018 - Proceedings of the
  World Wide Web Conference, WWW 2018}} (\bibinfo{year}{2018}),
  \bibinfo{pages}{823--831}.
\newblock
\showISBNx{9781450356398}
\urldef\tempurl%
\url{https://doi.org/10.1145/3178876.3186130}
\showDOI{\tempurl}
\showeprint[arxiv]{1803.01731}


\bibitem[\protect\citeauthoryear{Google}{Google}{[n.d.]}]%
        {Trending70online}
\bibfield{author}{\bibinfo{person}{Google}.} \bibinfo{year}{[n.d.]}\natexlab{}.
\newblock \bibinfo{title}{Trending on YouTube - YouTube Help}.
\newblock
  \bibinfo{howpublished}{\url{https://support.google.com/youtube/answer/7239739?hl=en}}.
\newblock
\newblock
\shownote{(Accessed on 09/09/2021).}


\bibitem[\protect\citeauthoryear{Govaerts, Verbert, Duval, and Pardo}{Govaerts
  et~al\mbox{.}}{2012}]%
        {govaerts2012student}
\bibfield{author}{\bibinfo{person}{Sten Govaerts}, \bibinfo{person}{Katrien
  Verbert}, \bibinfo{person}{Erik Duval}, {and} \bibinfo{person}{Abelardo
  Pardo}.} \bibinfo{year}{2012}\natexlab{}.
\newblock \showarticletitle{The student activity meter for awareness and
  self-reflection}.
\newblock In \bibinfo{booktitle}{\emph{CHI'12 Extended Abstracts on Human
  Factors in Computing Systems}}. \bibinfo{publisher}{Association for Computing
  Machinery}, \bibinfo{address}{New York, NY, USA}, \bibinfo{pages}{869--884}.
\newblock


\bibitem[\protect\citeauthoryear{Grant, Franklin, and Langford}{Grant
  et~al\mbox{.}}{2002}]%
        {Grant2002}
\bibfield{author}{\bibinfo{person}{Anthony~M. Grant}, \bibinfo{person}{John
  Franklin}, {and} \bibinfo{person}{Peter Langford}.}
  \bibinfo{year}{2002}\natexlab{}.
\newblock \showarticletitle{{The self-reflection and insight scale: A new
  measure of private self-consciousness}}.
\newblock \bibinfo{journal}{\emph{Social Behavior and Personality}}
  \bibinfo{volume}{30}, \bibinfo{number}{8} (\bibinfo{year}{2002}),
  \bibinfo{pages}{821--836}.
\newblock
\showISSN{03012212}
\urldef\tempurl%
\url{https://doi.org/10.2224/sbp.2002.30.8.821}
\showDOI{\tempurl}


\bibitem[\protect\citeauthoryear{Grevet and Gilbert}{Grevet and
  Gilbert}{2015}]%
        {grevet2015piggyback}
\bibfield{author}{\bibinfo{person}{Catherine Grevet} {and}
  \bibinfo{person}{Eric Gilbert}.} \bibinfo{year}{2015}\natexlab{}.
\newblock \showarticletitle{Piggyback prototyping: Using existing, large-scale
  social computing systems to prototype new ones}. In
  \bibinfo{booktitle}{\emph{Proceedings of the 33rd Annual ACM Conference on
  Human Factors in Computing Systems}}. \bibinfo{publisher}{Association for
  Computing Machinery}, \bibinfo{address}{New York, NY, USA},
  \bibinfo{pages}{4047--4056}.
\newblock


\bibitem[\protect\citeauthoryear{Grimes, Landry, and Grinter}{Grimes
  et~al\mbox{.}}{2010}]%
        {grimes2010characteristics}
\bibfield{author}{\bibinfo{person}{Andrea Grimes}, \bibinfo{person}{Brian~M
  Landry}, {and} \bibinfo{person}{Rebecca~E Grinter}.}
  \bibinfo{year}{2010}\natexlab{}.
\newblock \showarticletitle{Characteristics of shared health reflections in a
  local community}. In \bibinfo{booktitle}{\emph{Proceedings of the 2010 ACM
  conference on Computer supported cooperative work}}.
  \bibinfo{publisher}{Association for Computing Machinery},
  \bibinfo{address}{New York, NY, USA}, \bibinfo{pages}{435--444}.
\newblock


\bibitem[\protect\citeauthoryear{Gunawardana and Shani}{Gunawardana and
  Shani}{2015}]%
        {gunawardana2015evaluating}
\bibfield{author}{\bibinfo{person}{Asela Gunawardana} {and}
  \bibinfo{person}{Guy Shani}.} \bibinfo{year}{2015}\natexlab{}.
\newblock \bibinfo{booktitle}{\emph{Evaluating Recommender Systems}}.
\newblock \bibinfo{publisher}{Springer US}, \bibinfo{address}{Boston, MA},
  \bibinfo{pages}{265--308}.
\newblock
\showISBNx{978-1-4899-7637-6}
\urldef\tempurl%
\url{https://doi.org/10.1007/978-1-4899-7637-6_8}
\showDOI{\tempurl}


\bibitem[\protect\citeauthoryear{Halttu and Oinas-kukkonen}{Halttu and
  Oinas-kukkonen}{2017}]%
        {Halttu2017}
\bibfield{author}{\bibinfo{person}{Kirsi Halttu} {and} \bibinfo{person}{Harri
  Oinas-kukkonen}.} \bibinfo{year}{2017}\natexlab{}.
\newblock \showarticletitle{{Human – Computer Interaction Persuading to
  Reflect : Role of Reflection and Insight in Persuasive Systems Design for
  Physical Health Persuading to Reflect : Role of Reflection and Insight in
  Persuasive Systems Design for Physical Health}}.
\newblock \bibinfo{journal}{\emph{Human–Computer Interaction}}
  \bibinfo{volume}{32}, \bibinfo{number}{5-6} (\bibinfo{year}{2017}),
  \bibinfo{pages}{381--412}.
\newblock
\showISSN{0737-0024}
\urldef\tempurl%
\url{https://doi.org/10.1080/07370024.2017.1283227}
\showDOI{\tempurl}


\bibitem[\protect\citeauthoryear{Hedeker, du~Toit, Demirtas, and
  Gibbons}{Hedeker et~al\mbox{.}}{2018}]%
        {hedeker2018note}
\bibfield{author}{\bibinfo{person}{Donald Hedeker}, \bibinfo{person}{Stephen~HC
  du Toit}, \bibinfo{person}{Hakan Demirtas}, {and} \bibinfo{person}{Robert~D
  Gibbons}.} \bibinfo{year}{2018}\natexlab{}.
\newblock \showarticletitle{A note on marginalization of regression parameters
  from mixed models of binary outcomes}.
\newblock \bibinfo{journal}{\emph{Biometrics}} \bibinfo{volume}{74},
  \bibinfo{number}{1} (\bibinfo{year}{2018}), \bibinfo{pages}{354--361}.
\newblock


\bibitem[\protect\citeauthoryear{Helberger, Karppinen, and
  D’acunto}{Helberger et~al\mbox{.}}{2018}]%
        {helberger2018exposure}
\bibfield{author}{\bibinfo{person}{Natali Helberger}, \bibinfo{person}{Kari
  Karppinen}, {and} \bibinfo{person}{Lucia D’acunto}.}
  \bibinfo{year}{2018}\natexlab{}.
\newblock \showarticletitle{Exposure diversity as a design principle for
  recommender systems}.
\newblock \bibinfo{journal}{\emph{Information, Communication \& Society}}
  \bibinfo{volume}{21}, \bibinfo{number}{2} (\bibinfo{year}{2018}),
  \bibinfo{pages}{191--207}.
\newblock


\bibitem[\protect\citeauthoryear{Hijikata, Shimizu, and Nishida}{Hijikata
  et~al\mbox{.}}{2009}]%
        {hijikata2009discovery}
\bibfield{author}{\bibinfo{person}{Yoshinori Hijikata}, \bibinfo{person}{Takuya
  Shimizu}, {and} \bibinfo{person}{Shogo Nishida}.}
  \bibinfo{year}{2009}\natexlab{}.
\newblock \showarticletitle{Discovery-oriented collaborative filtering for
  improving user satisfaction}. In \bibinfo{booktitle}{\emph{Proceedings of the
  14th international conference on Intelligent user interfaces}}.
  \bibinfo{publisher}{Association for Computing Machinery},
  \bibinfo{address}{New York, NY, USA}, \bibinfo{pages}{67--76}.
\newblock


\bibitem[\protect\citeauthoryear{Hughes, Varma, Pettigrew, and Albert}{Hughes
  et~al\mbox{.}}{2017}]%
        {hughes2017african}
\bibfield{author}{\bibinfo{person}{Travonia~B Hughes}, \bibinfo{person}{Vijay~R
  Varma}, \bibinfo{person}{Corinne Pettigrew}, {and} \bibinfo{person}{Marilyn~S
  Albert}.} \bibinfo{year}{2017}\natexlab{}.
\newblock \showarticletitle{African Americans and clinical research: evidence
  concerning barriers and facilitators to participation and recruitment
  recommendations}.
\newblock \bibinfo{journal}{\emph{The Gerontologist}} \bibinfo{volume}{57},
  \bibinfo{number}{2} (\bibinfo{year}{2017}), \bibinfo{pages}{348--358}.
\newblock


\bibitem[\protect\citeauthoryear{Johnston, Amitani, and Edmonds}{Johnston
  et~al\mbox{.}}{2005}]%
        {johnston2005amplifying}
\bibfield{author}{\bibinfo{person}{Andrew Johnston}, \bibinfo{person}{Shigeki
  Amitani}, {and} \bibinfo{person}{Ernest Edmonds}.}
  \bibinfo{year}{2005}\natexlab{}.
\newblock \showarticletitle{Amplifying reflective thinking in musical
  performance}. In \bibinfo{booktitle}{\emph{Proceedings of the 5th conference
  on Creativity \& cognition}}. \bibinfo{publisher}{Association for Computing
  Machinery}, \bibinfo{address}{New York, NY, USA}, \bibinfo{pages}{166--175}.
\newblock


\bibitem[\protect\citeauthoryear{Keen}{Keen}{2021}]%
        {KeenExpa82online}
\bibfield{author}{\bibinfo{person}{Keen}.} \bibinfo{year}{2021}\natexlab{}.
\newblock \bibinfo{title}{Keen | Expand your interests}.
\newblock \bibinfo{howpublished}{\url{https://staykeen.com/about}}.
\newblock
\newblock
\shownote{(Accessed on 09/07/2021).}


\bibitem[\protect\citeauthoryear{Kiskola, Olsson, V\"{a}\"{a}t\"{a}j\"{a},
  H.~Syrj\"{a}m\"{a}ki, Rantasila, Isokoski, Ilves, and Surakka}{Kiskola
  et~al\mbox{.}}{2021}]%
        {kiskola2021applying}
\bibfield{author}{\bibinfo{person}{Joel Kiskola}, \bibinfo{person}{Thomas
  Olsson}, \bibinfo{person}{Heli V\"{a}\"{a}t\"{a}j\"{a}},
  \bibinfo{person}{Aleksi H.~Syrj\"{a}m\"{a}ki}, \bibinfo{person}{Anna
  Rantasila}, \bibinfo{person}{Poika Isokoski}, \bibinfo{person}{Mirja Ilves},
  {and} \bibinfo{person}{Veikko Surakka}.} \bibinfo{year}{2021}\natexlab{}.
\newblock \bibinfo{booktitle}{\emph{Applying Critical Voice in Design of User
  Interfaces for Supporting Self-Reflection and Emotion Regulation in Online
  News Commenting}}.
\newblock \bibinfo{publisher}{Association for Computing Machinery},
  \bibinfo{address}{New York, NY, USA}.
\newblock
\showISBNx{9781450380966}
\urldef\tempurl%
\url{https://doi.org/10.1145/3411764.3445783}
\showURL{%
\tempurl}


\bibitem[\protect\citeauthoryear{Kontiza, Loboda, Deladiennee, Castagnos, and
  Naudet}{Kontiza et~al\mbox{.}}{2018}]%
        {kontiza2018museum}
\bibfield{author}{\bibinfo{person}{Kalliopi Kontiza}, \bibinfo{person}{Olga
  Loboda}, \bibinfo{person}{Louis Deladiennee}, \bibinfo{person}{Sylvain
  Castagnos}, {and} \bibinfo{person}{Yannick Naudet}.}
  \bibinfo{year}{2018}\natexlab{}.
\newblock \showarticletitle{A museum app to trigger users' reflection}. In
  \bibinfo{booktitle}{\emph{International Workshop on Mobile Access to Cultural
  Heritage (MobileCH2018)}}. \bibinfo{address}{Barcelona, Spain}.
\newblock


\bibitem[\protect\citeauthoryear{Koo and Skinner}{Koo and Skinner}{2005}]%
        {koo2005challenges}
\bibfield{author}{\bibinfo{person}{Malcolm Koo} {and} \bibinfo{person}{Harvey
  Skinner}.} \bibinfo{year}{2005}\natexlab{}.
\newblock \showarticletitle{Challenges of internet recruitment: a case study
  with disappointing results}.
\newblock \bibinfo{journal}{\emph{Journal of Medical Internet Research}}
  \bibinfo{volume}{7}, \bibinfo{number}{1} (\bibinfo{year}{2005}),
  \bibinfo{pages}{e6}.
\newblock


\bibitem[\protect\citeauthoryear{Korinek, Phatak, Martin, Freigoun, Rivera,
  Adams, Klasnja, Buman, and Hekler}{Korinek et~al\mbox{.}}{2018}]%
        {korinek2018adaptive}
\bibfield{author}{\bibinfo{person}{Elizabeth~V Korinek},
  \bibinfo{person}{Sayali~S Phatak}, \bibinfo{person}{Cesar~A Martin},
  \bibinfo{person}{Mohammad~T Freigoun}, \bibinfo{person}{Daniel~E Rivera},
  \bibinfo{person}{Marc~A Adams}, \bibinfo{person}{Pedja Klasnja},
  \bibinfo{person}{Matthew~P Buman}, {and} \bibinfo{person}{Eric~B Hekler}.}
  \bibinfo{year}{2018}\natexlab{}.
\newblock \showarticletitle{Adaptive step goals and rewards: a longitudinal
  growth model of daily steps for a smartphone-based walking intervention}.
\newblock \bibinfo{journal}{\emph{Journal of behavioral medicine}}
  \bibinfo{volume}{41}, \bibinfo{number}{1} (\bibinfo{year}{2018}),
  \bibinfo{pages}{74--86}.
\newblock


\bibitem[\protect\citeauthoryear{Lab}{Lab}{2017}]%
        {Overview78online}
\bibfield{author}{\bibinfo{person}{MIT~Media Lab}.}
  \bibinfo{year}{2017}\natexlab{}.
\newblock \bibinfo{title}{Overview ‹ FlipFeed — MIT Media Lab}.
\newblock
  \bibinfo{howpublished}{\url{https://www.media.mit.edu/projects/flipfeed/overview/}}.
\newblock
\newblock
\shownote{(Accessed on 08/31/2021).}


\bibitem[\protect\citeauthoryear{Lamberty and Kolodner}{Lamberty and
  Kolodner}{2005}]%
        {lamberty2005camera}
\bibfield{author}{\bibinfo{person}{KK Lamberty} {and} \bibinfo{person}{Janet~L
  Kolodner}.} \bibinfo{year}{2005}\natexlab{}.
\newblock \showarticletitle{Camera talk: Making the camera a partial
  participant}. In \bibinfo{booktitle}{\emph{Proceedings of the SIGCHI
  Conference on Human Factors in Computing Systems}}.
  \bibinfo{publisher}{Association for Computing Machinery},
  \bibinfo{address}{New York, NY, USA}, \bibinfo{pages}{839--848}.
\newblock


\bibitem[\protect\citeauthoryear{Lathia, Hailes, Capra, and Amatriain}{Lathia
  et~al\mbox{.}}{2010}]%
        {lathia2010temporal}
\bibfield{author}{\bibinfo{person}{Neal Lathia}, \bibinfo{person}{Stephen
  Hailes}, \bibinfo{person}{Licia Capra}, {and} \bibinfo{person}{Xavier
  Amatriain}.} \bibinfo{year}{2010}\natexlab{}.
\newblock \showarticletitle{Temporal diversity in recommender systems}. In
  \bibinfo{booktitle}{\emph{Proceedings of the 33rd international ACM SIGIR
  conference on Research and development in information retrieval}}.
  \bibinfo{publisher}{Association for Computing Machinery},
  \bibinfo{address}{New York, NY, USA}, \bibinfo{pages}{210--217}.
\newblock


\bibitem[\protect\citeauthoryear{Lee and Dey}{Lee and Dey}{2011}]%
        {lee2011reflecting}
\bibfield{author}{\bibinfo{person}{Matthew~L Lee} {and}
  \bibinfo{person}{Anind~K Dey}.} \bibinfo{year}{2011}\natexlab{}.
\newblock \showarticletitle{Reflecting on pills and phone use: supporting
  awareness of functional abilities for older adults}. In
  \bibinfo{booktitle}{\emph{Proceedings of the SIGCHI Conference on Human
  Factors in Computing Systems}}. \bibinfo{publisher}{Association for Computing
  Machinery}, \bibinfo{address}{New York, NY, USA},
  \bibinfo{pages}{2095--2104}.
\newblock


\bibitem[\protect\citeauthoryear{Li, Dey, and Forlizzi}{Li
  et~al\mbox{.}}{2009}]%
        {li2009grafitter}
\bibfield{author}{\bibinfo{person}{Ian Li}, \bibinfo{person}{Anind Dey}, {and}
  \bibinfo{person}{Jodi Forlizzi}.} \bibinfo{year}{2009}\natexlab{}.
\newblock \showarticletitle{Grafitter: leveraging social media for self
  reflection}.
\newblock \bibinfo{journal}{\emph{XRDS: Crossroads, The ACM Magazine for
  Students}} \bibinfo{volume}{16}, \bibinfo{number}{2} (\bibinfo{year}{2009}),
  \bibinfo{pages}{12--13}.
\newblock


\bibitem[\protect\citeauthoryear{Li, Dey, and Forlizzi}{Li
  et~al\mbox{.}}{2010}]%
        {li2010stage}
\bibfield{author}{\bibinfo{person}{Ian Li}, \bibinfo{person}{Anind Dey}, {and}
  \bibinfo{person}{Jodi Forlizzi}.} \bibinfo{year}{2010}\natexlab{}.
\newblock \showarticletitle{A stage-based model of personal informatics
  systems}. In \bibinfo{booktitle}{\emph{Proceedings of the SIGCHI conference
  on human factors in computing systems}}. \bibinfo{publisher}{Association for
  Computing Machinery}, \bibinfo{address}{New York, NY, USA},
  \bibinfo{pages}{557--566}.
\newblock


\bibitem[\protect\citeauthoryear{Li, Dey, and Forlizzi}{Li
  et~al\mbox{.}}{2011}]%
        {li2011understanding}
\bibfield{author}{\bibinfo{person}{Ian Li}, \bibinfo{person}{Anind~K Dey},
  {and} \bibinfo{person}{Jodi Forlizzi}.} \bibinfo{year}{2011}\natexlab{}.
\newblock \showarticletitle{Understanding my data, myself: supporting
  self-reflection with ubicomp technologies}. In
  \bibinfo{booktitle}{\emph{Proceedings of the 13th international conference on
  Ubiquitous computing}}. \bibinfo{publisher}{Association for Computing
  Machinery}, \bibinfo{address}{New York, NY, USA}, \bibinfo{pages}{405--414}.
\newblock


\bibitem[\protect\citeauthoryear{Lin, Mamykina, Lindtner, Delajoux, and
  Strub}{Lin et~al\mbox{.}}{2006}]%
        {lin2006fish}
\bibfield{author}{\bibinfo{person}{James~J Lin}, \bibinfo{person}{Lena
  Mamykina}, \bibinfo{person}{Silvia Lindtner}, \bibinfo{person}{Gregory
  Delajoux}, {and} \bibinfo{person}{Henry~B Strub}.}
  \bibinfo{year}{2006}\natexlab{}.
\newblock \showarticletitle{Fish’n’Steps: Encouraging physical activity
  with an interactive computer game}. In
  \bibinfo{booktitle}{\emph{International conference on ubiquitous computing}}.
  Springer, \bibinfo{publisher}{Springer Berlin Heidelberg},
  \bibinfo{address}{Berlin, Heidelberg}, \bibinfo{pages}{261--278}.
\newblock


\bibitem[\protect\citeauthoryear{Malacria, Scarr, Cockburn, Gutwin, and
  Grossman}{Malacria et~al\mbox{.}}{2013}]%
        {malacria2013skillometers}
\bibfield{author}{\bibinfo{person}{Sylvain Malacria}, \bibinfo{person}{Joey
  Scarr}, \bibinfo{person}{Andy Cockburn}, \bibinfo{person}{Carl Gutwin}, {and}
  \bibinfo{person}{Tovi Grossman}.} \bibinfo{year}{2013}\natexlab{}.
\newblock \showarticletitle{Skillometers: Reflective widgets that motivate and
  help users to improve performance}. In \bibinfo{booktitle}{\emph{Proceedings
  of the 26th annual ACM symposium on User interface software and technology}}.
  \bibinfo{publisher}{Association for Computing Machinery},
  \bibinfo{address}{New York, NY, USA}, \bibinfo{pages}{321--330}.
\newblock


\bibitem[\protect\citeauthoryear{McNee, Riedl, and Konstan}{McNee
  et~al\mbox{.}}{2006}]%
        {mcnee2006being}
\bibfield{author}{\bibinfo{person}{Sean~M McNee}, \bibinfo{person}{John Riedl},
  {and} \bibinfo{person}{Joseph~A Konstan}.} \bibinfo{year}{2006}\natexlab{}.
\newblock \showarticletitle{Being accurate is not enough: how accuracy metrics
  have hurt recommender systems}. In \bibinfo{booktitle}{\emph{CHI'06 extended
  abstracts on Human factors in computing systems}}.
  \bibinfo{publisher}{Association for Computing Machinery},
  \bibinfo{address}{New York, NY, USA}, \bibinfo{pages}{1097--1101}.
\newblock


\bibitem[\protect\citeauthoryear{Medrek}{Medrek}{2018}]%
        {Medrek2018}
\bibfield{author}{\bibinfo{person}{Ania Medrek}.}
  \bibinfo{year}{2018}\natexlab{}.
\newblock \showarticletitle{{NEWS BY ASSOCIATION: Designing a way out of the
  echo chamber}}.
\newblock  \bibinfo{number}{April} (\bibinfo{year}{2018}).
\newblock


\bibitem[\protect\citeauthoryear{Michie, Richardson, Johnston, Abraham,
  Francis, Hardeman, Eccles, Cane, and Wood}{Michie et~al\mbox{.}}{2013}]%
        {michie2013behavior}
\bibfield{author}{\bibinfo{person}{Susan Michie}, \bibinfo{person}{Michelle
  Richardson}, \bibinfo{person}{Marie Johnston}, \bibinfo{person}{Charles
  Abraham}, \bibinfo{person}{Jill Francis}, \bibinfo{person}{Wendy Hardeman},
  \bibinfo{person}{Martin~P Eccles}, \bibinfo{person}{James Cane}, {and}
  \bibinfo{person}{Caroline~E Wood}.} \bibinfo{year}{2013}\natexlab{}.
\newblock \showarticletitle{The behavior change technique taxonomy (v1) of 93
  hierarchically clustered techniques: building an international consensus for
  the reporting of behavior change interventions}.
\newblock \bibinfo{journal}{\emph{Annals of behavioral medicine}}
  \bibinfo{volume}{46}, \bibinfo{number}{1} (\bibinfo{year}{2013}),
  \bibinfo{pages}{81--95}.
\newblock


\bibitem[\protect\citeauthoryear{Mols, van~den Hoven, and Eggen}{Mols
  et~al\mbox{.}}{2016}]%
        {Mols2016}
\bibfield{author}{\bibinfo{person}{Ine Mols}, \bibinfo{person}{Elise van~den
  Hoven}, {and} \bibinfo{person}{Berry Eggen}.}
  \bibinfo{year}{2016}\natexlab{}.
\newblock \showarticletitle{Informing Design for Reflection: An Overview of
  Current Everyday Practices}. In \bibinfo{booktitle}{\emph{Proceedings of the
  9th Nordic Conference on Human-Computer Interaction}} (Gothenburg, Sweden)
  \emph{(\bibinfo{series}{NordiCHI '16})}. \bibinfo{publisher}{Association for
  Computing Machinery}, \bibinfo{address}{New York, NY, USA}, Article
  \bibinfo{articleno}{21}, \bibinfo{numpages}{10}~pages.
\newblock
\showISBNx{9781450347631}
\urldef\tempurl%
\url{https://doi.org/10.1145/2971485.2971494}
\showDOI{\tempurl}


\bibitem[\protect\citeauthoryear{Munson, Lee, and Resnick}{Munson
  et~al\mbox{.}}{2013b}]%
        {munson2013encouraging}
\bibfield{author}{\bibinfo{person}{Sean Munson}, \bibinfo{person}{Stephanie
  Lee}, {and} \bibinfo{person}{Paul Resnick}.}
  \bibinfo{year}{2013}\natexlab{b}.
\newblock \showarticletitle{Encouraging reading of diverse political viewpoints
  with a browser widget}. In \bibinfo{booktitle}{\emph{Proceedings of The
  International AAAI Conference on Web and Social Media}},
  Vol.~\bibinfo{volume}{7}.
\newblock


\bibitem[\protect\citeauthoryear{Munson, Cavusoglu, Frisch, and Fels}{Munson
  et~al\mbox{.}}{2013a}]%
        {munson2013sociotechnical}
\bibfield{author}{\bibinfo{person}{Sean~A Munson}, \bibinfo{person}{Hasan
  Cavusoglu}, \bibinfo{person}{Larry Frisch}, {and} \bibinfo{person}{Sidney
  Fels}.} \bibinfo{year}{2013}\natexlab{a}.
\newblock \showarticletitle{Sociotechnical challenges and progress in using
  social media for health}.
\newblock \bibinfo{journal}{\emph{Journal of medical Internet research}}
  \bibinfo{volume}{15}, \bibinfo{number}{10} (\bibinfo{year}{2013}),
  \bibinfo{pages}{e226}.
\newblock


\bibitem[\protect\citeauthoryear{Napoli}{Napoli}{2011}]%
        {napoli2011exposure}
\bibfield{author}{\bibinfo{person}{Philip~M Napoli}.}
  \bibinfo{year}{2011}\natexlab{}.
\newblock \showarticletitle{Exposure diversity reconsidered}.
\newblock \bibinfo{journal}{\emph{Journal of information policy}}
  \bibinfo{volume}{1} (\bibinfo{year}{2011}), \bibinfo{pages}{246--259}.
\newblock


\bibitem[\protect\citeauthoryear{Nguyen, Hui, Harper, Terveen, and
  Konstan}{Nguyen et~al\mbox{.}}{2014}]%
        {nguyen2014exploring}
\bibfield{author}{\bibinfo{person}{Tien~T Nguyen}, \bibinfo{person}{Pik-Mai
  Hui}, \bibinfo{person}{F~Maxwell Harper}, \bibinfo{person}{Loren Terveen},
  {and} \bibinfo{person}{Joseph~A Konstan}.} \bibinfo{year}{2014}\natexlab{}.
\newblock \showarticletitle{Exploring the filter bubble: the effect of using
  recommender systems on content diversity}. In
  \bibinfo{booktitle}{\emph{Proceedings of the 23rd international conference on
  World wide web}}. \bibinfo{pages}{677--686}.
\newblock


\bibitem[\protect\citeauthoryear{Nichols and Kang}{Nichols and Kang}{2012}]%
        {nichols2012asking}
\bibfield{author}{\bibinfo{person}{Jeffrey Nichols} {and}
  \bibinfo{person}{Jeon-Hyung Kang}.} \bibinfo{year}{2012}\natexlab{}.
\newblock \showarticletitle{Asking questions of targeted strangers on social
  networks}. In \bibinfo{booktitle}{\emph{Proceedings of the ACM 2012
  conference on Computer Supported Cooperative Work}}.
  \bibinfo{publisher}{Association for Computing Machinery},
  \bibinfo{address}{New York, NY, USA}, \bibinfo{pages}{999--1002}.
\newblock


\bibitem[\protect\citeauthoryear{Niess and Wo{\'z}niak}{Niess and
  Wo{\'z}niak}{2018}]%
        {niess2018supporting}
\bibfield{author}{\bibinfo{person}{Jasmin Niess} {and}
  \bibinfo{person}{Pawe{\l}~W Wo{\'z}niak}.} \bibinfo{year}{2018}\natexlab{}.
\newblock \showarticletitle{Supporting meaningful personal fitness: The tracker
  goal evolution model}. In \bibinfo{booktitle}{\emph{Proceedings of the 2018
  CHI Conference on Human Factors in Computing Systems}}.
  \bibinfo{publisher}{Association for Computing Machinery},
  \bibinfo{address}{New York, NY, USA}, \bibinfo{pages}{1--12}.
\newblock


\bibitem[\protect\citeauthoryear{Nussbaumer, Kravcik, and Albert}{Nussbaumer
  et~al\mbox{.}}{2012}]%
        {nussbaumer2012supporting}
\bibfield{author}{\bibinfo{person}{Alexander Nussbaumer},
  \bibinfo{person}{Milos Kravcik}, {and} \bibinfo{person}{Dietrich Albert}.}
  \bibinfo{year}{2012}\natexlab{}.
\newblock \showarticletitle{Supporting self-reflection in personal learning
  environments through user feedback.}. In \bibinfo{booktitle}{\emph{UMAP
  Workshops}}.
\newblock


\bibitem[\protect\citeauthoryear{Obadimu, Mead, Hussain, and Agarwal}{Obadimu
  et~al\mbox{.}}{2019}]%
        {obadimu2019identifying}
\bibfield{author}{\bibinfo{person}{Adewale Obadimu}, \bibinfo{person}{Esther
  Mead}, \bibinfo{person}{Muhammad~Nihal Hussain}, {and} \bibinfo{person}{Nitin
  Agarwal}.} \bibinfo{year}{2019}\natexlab{}.
\newblock \showarticletitle{Identifying toxicity within youtube video comment}.
  In \bibinfo{booktitle}{\emph{International Conference on Social Computing,
  Behavioral-Cultural Modeling and Prediction and Behavior Representation in
  Modeling and Simulation}}. Springer, \bibinfo{pages}{214--223}.
\newblock


\bibitem[\protect\citeauthoryear{Ookalkar, Reddy, and Gilbert}{Ookalkar
  et~al\mbox{.}}{2019a}]%
        {Ookalkar2019}
\bibfield{author}{\bibinfo{person}{Ruchi Ookalkar},
  \bibinfo{person}{Kolli~Vishal Reddy}, {and} \bibinfo{person}{Eric Gilbert}.}
  \bibinfo{year}{2019}\natexlab{a}.
\newblock \showarticletitle{{Pop: Bursting news filter bubbles on twiter
  through diverse exposure}}.
\newblock \bibinfo{journal}{\emph{Proceedings of the ACM Conference on Computer
  Supported Cooperative Work, CSCW}} (\bibinfo{year}{2019}),
  \bibinfo{pages}{18--21}.
\newblock
\showISBNx{9781450366922}
\urldef\tempurl%
\url{https://doi.org/10.1145/3311957.3359513}
\showDOI{\tempurl}


\bibitem[\protect\citeauthoryear{Ookalkar, Reddy, and Gilbert}{Ookalkar
  et~al\mbox{.}}{2019b}]%
        {ookalkar2019pop}
\bibfield{author}{\bibinfo{person}{Ruchi Ookalkar},
  \bibinfo{person}{Kolli~Vishal Reddy}, {and} \bibinfo{person}{Eric Gilbert}.}
  \bibinfo{year}{2019}\natexlab{b}.
\newblock \showarticletitle{Pop: Bursting News Filter Bubbles on Twitter
  Through Diverse Exposure}. In \bibinfo{booktitle}{\emph{Conference Companion
  Publication of the 2019 on Computer Supported Cooperative Work and Social
  Computing}}. \bibinfo{pages}{18--22}.
\newblock


\bibitem[\protect\citeauthoryear{Pariser}{Pariser}{2011}]%
        {pariser2011filter}
\bibfield{author}{\bibinfo{person}{Eli Pariser}.}
  \bibinfo{year}{2011}\natexlab{}.
\newblock \bibinfo{booktitle}{\emph{The filter bubble: How the new personalized
  web is changing what we read and how we think}}.
\newblock \bibinfo{publisher}{Penguin}.
\newblock


\bibitem[\protect\citeauthoryear{Ploderer, Reitberger, Oinas-Kukkonen, and van
  Gemert-Pijnen}{Ploderer et~al\mbox{.}}{2014}]%
        {ploderer2014social}
\bibfield{author}{\bibinfo{person}{Bernd Ploderer}, \bibinfo{person}{Wolfgang
  Reitberger}, \bibinfo{person}{Harri Oinas-Kukkonen}, {and}
  \bibinfo{person}{Julia van Gemert-Pijnen}.} \bibinfo{year}{2014}\natexlab{}.
\newblock \bibinfo{title}{Social interaction and reflection for behaviour
  change}.
\newblock
\newblock


\bibitem[\protect\citeauthoryear{Ramo and Prochaska}{Ramo and
  Prochaska}{2012}]%
        {ramo2012broad}
\bibfield{author}{\bibinfo{person}{Danielle~E Ramo} {and}
  \bibinfo{person}{Judith~J Prochaska}.} \bibinfo{year}{2012}\natexlab{}.
\newblock \showarticletitle{Broad reach and targeted recruitment using Facebook
  for an online survey of young adult substance use}.
\newblock \bibinfo{journal}{\emph{Journal of medical Internet research}}
  \bibinfo{volume}{14}, \bibinfo{number}{1} (\bibinfo{year}{2012}),
  \bibinfo{pages}{e28}.
\newblock


\bibitem[\protect\citeauthoryear{Resnick, Garrett, Kriplean, Munson, and
  Stroud}{Resnick et~al\mbox{.}}{2013}]%
        {resnick2013bursting}
\bibfield{author}{\bibinfo{person}{Paul Resnick}, \bibinfo{person}{R~Kelly
  Garrett}, \bibinfo{person}{Travis Kriplean}, \bibinfo{person}{Sean~A Munson},
  {and} \bibinfo{person}{Natalie~Jomini Stroud}.}
  \bibinfo{year}{2013}\natexlab{}.
\newblock \showarticletitle{Bursting your (filter) bubble: strategies for
  promoting diverse exposure}. In \bibinfo{booktitle}{\emph{Proceedings of the
  2013 conference on Computer supported cooperative work companion}}.
  \bibinfo{publisher}{Association for Computing Machinery},
  \bibinfo{address}{New York, NY, USA}, \bibinfo{pages}{95--100}.
\newblock


\bibitem[\protect\citeauthoryear{Rizopoulos}{Rizopoulos}{2019}]%
        {rizopoulos2019glmmadaptive}
\bibfield{author}{\bibinfo{person}{Dimitris Rizopoulos}.}
  \bibinfo{year}{2019}\natexlab{}.
\newblock \showarticletitle{GLMMadaptive: generalized linear mixed models using
  adaptive Gaussian quadrature}.
\newblock \bibinfo{journal}{\emph{R package version 0.5--1}}
  (\bibinfo{year}{2019}).
\newblock


\bibitem[\protect\citeauthoryear{Sas and Dix}{Sas and Dix}{2011}]%
        {sas2011designing}
\bibfield{author}{\bibinfo{person}{Corina Sas} {and} \bibinfo{person}{Alan
  Dix}.} \bibinfo{year}{2011}\natexlab{}.
\newblock \showarticletitle{Designing for reflection on personal experience}.
\newblock \bibinfo{journal}{\emph{International Journal of Human-Computer
  Studies}} \bibinfo{volume}{69}, \bibinfo{number}{5} (\bibinfo{year}{2011}),
  \bibinfo{pages}{281--282}.
\newblock


\bibitem[\protect\citeauthoryear{Schaap}{Schaap}{2020}]%
        {Schaap2020}
\bibfield{author}{\bibinfo{person}{Jorrit Schaap}.}
  \bibinfo{year}{2020}\natexlab{}.
\newblock \showarticletitle{{Bubble Trouble – Venture Out of Your Filter
  Bubbles}}.
\newblock  (\bibinfo{year}{2020}), \bibinfo{pages}{1--14}.
\newblock


\bibitem[\protect\citeauthoryear{Seidman}{Seidman}{2013}]%
        {seidman2013self}
\bibfield{author}{\bibinfo{person}{Gwendolyn Seidman}.}
  \bibinfo{year}{2013}\natexlab{}.
\newblock \showarticletitle{Self-presentation and belonging on Facebook: How
  personality influences social media use and motivations}.
\newblock \bibinfo{journal}{\emph{Personality and individual differences}}
  \bibinfo{volume}{54}, \bibinfo{number}{3} (\bibinfo{year}{2013}),
  \bibinfo{pages}{402--407}.
\newblock


\bibitem[\protect\citeauthoryear{Sheth, Bell, Arora, and Kaiser}{Sheth
  et~al\mbox{.}}{2011}]%
        {sheth2011towards}
\bibfield{author}{\bibinfo{person}{Swapneel~Kalpesh Sheth},
  \bibinfo{person}{Jonathan~Schaffer Bell}, \bibinfo{person}{Nipun Arora},
  {and} \bibinfo{person}{Gail~E Kaiser}.} \bibinfo{year}{2011}\natexlab{}.
\newblock \showarticletitle{Towards diversity in recommendations using social
  networks}.
\newblock  (\bibinfo{year}{2011}).
\newblock


\bibitem[\protect\citeauthoryear{Stecula and Pickup}{Stecula and
  Pickup}{2021}]%
        {stecula2021social}
\bibfield{author}{\bibinfo{person}{Dominik~A Stecula} {and}
  \bibinfo{person}{Mark Pickup}.} \bibinfo{year}{2021}\natexlab{}.
\newblock \showarticletitle{Social media, cognitive reflection, and conspiracy
  beliefs}.
\newblock \bibinfo{journal}{\emph{Frontiers in Political Science}}
  \bibinfo{volume}{3} (\bibinfo{year}{2021}), \bibinfo{pages}{62}.
\newblock


\bibitem[\protect\citeauthoryear{Thieme, Wallace, Johnson, McCarthy, Lindley,
  Wright, Olivier, and Meyer}{Thieme et~al\mbox{.}}{2013}]%
        {thieme2013design}
\bibfield{author}{\bibinfo{person}{Anja Thieme}, \bibinfo{person}{Jayne
  Wallace}, \bibinfo{person}{Paula Johnson}, \bibinfo{person}{John McCarthy},
  \bibinfo{person}{Si{\^a}n Lindley}, \bibinfo{person}{Peter Wright},
  \bibinfo{person}{Patrick Olivier}, {and} \bibinfo{person}{Thomas~D Meyer}.}
  \bibinfo{year}{2013}\natexlab{}.
\newblock \showarticletitle{Design to promote mindfulness practice and sense of
  self for vulnerable women in secure hospital services}. In
  \bibinfo{booktitle}{\emph{Proceedings of the SIGCHI Conference on Human
  Factors in Computing Systems}}. \bibinfo{publisher}{Association for Computing
  Machinery}, \bibinfo{address}{New York, NY, USA},
  \bibinfo{pages}{2647--2656}.
\newblock


\bibitem[\protect\citeauthoryear{Tseng and Bryant}{Tseng and Bryant}{2013}]%
        {tseng2013design}
\bibfield{author}{\bibinfo{person}{Tiffany Tseng} {and} \bibinfo{person}{Coram
  Bryant}.} \bibinfo{year}{2013}\natexlab{}.
\newblock \showarticletitle{Design, reflect, explore: encouraging children's
  reflections with mechanix}.
\newblock In \bibinfo{booktitle}{\emph{CHI'13 Extended Abstracts on Human
  Factors in Computing Systems}}. \bibinfo{publisher}{Association for Computing
  Machinery}, \bibinfo{address}{New York, NY, USA}, \bibinfo{pages}{619--624}.
\newblock


\bibitem[\protect\citeauthoryear{Valkanova, Jorda, Tomitsch, and
  Vande~Moere}{Valkanova et~al\mbox{.}}{2013}]%
        {valkanova2013reveal}
\bibfield{author}{\bibinfo{person}{Nina Valkanova}, \bibinfo{person}{Sergi
  Jorda}, \bibinfo{person}{Martin Tomitsch}, {and} \bibinfo{person}{Andrew
  Vande~Moere}.} \bibinfo{year}{2013}\natexlab{}.
\newblock \showarticletitle{Reveal-it! the impact of a social visualization
  projection on public awareness and discourse}. In
  \bibinfo{booktitle}{\emph{Proceedings of the SIGCHI Conference on Human
  Factors in Computing Systems}}. \bibinfo{publisher}{Association for Computing
  Machinery}, \bibinfo{address}{New York, NY, USA},
  \bibinfo{pages}{3461--3470}.
\newblock


\bibitem[\protect\citeauthoryear{Viera, Garrett, et~al\mbox{.}}{Viera
  et~al\mbox{.}}{2005}]%
        {viera2005understanding}
\bibfield{author}{\bibinfo{person}{Anthony~J Viera}, \bibinfo{person}{Joanne~M
  Garrett}, {et~al\mbox{.}}} \bibinfo{year}{2005}\natexlab{}.
\newblock \showarticletitle{Understanding interobserver agreement: the kappa
  statistic}.
\newblock \bibinfo{journal}{\emph{Fam med}} \bibinfo{volume}{37},
  \bibinfo{number}{5} (\bibinfo{year}{2005}), \bibinfo{pages}{360--363}.
\newblock


\bibitem[\protect\citeauthoryear{Wang and Yao}{Wang and Yao}{2020}]%
        {wang2020study}
\bibfield{author}{\bibinfo{person}{Yixue Wang} {and} \bibinfo{person}{Siyu
  Yao}.} \bibinfo{year}{2020}\natexlab{}.
\newblock \showarticletitle{STUDY ON INTENTION-AWARE RECOMMENDATION OF YOUTUBE
  VIDEOS}.
\newblock  (\bibinfo{year}{2020}).
\newblock


\bibitem[\protect\citeauthoryear{Wilhelm, Ramanathan, Bonomo, Jain, Chi, and
  Gillenwater}{Wilhelm et~al\mbox{.}}{2018}]%
        {Wilhelm2018}
\bibfield{author}{\bibinfo{person}{Mark Wilhelm}, \bibinfo{person}{Ajith
  Ramanathan}, \bibinfo{person}{Alexander Bonomo}, \bibinfo{person}{Sagar
  Jain}, \bibinfo{person}{Ed~H. Chi}, {and} \bibinfo{person}{Jennifer
  Gillenwater}.} \bibinfo{year}{2018}\natexlab{}.
\newblock \showarticletitle{Practical Diversified Recommendations on YouTube
  with Determinantal Point Processes}. In \bibinfo{booktitle}{\emph{Proceedings
  of the 27th ACM International Conference on Information and Knowledge
  Management}} (Torino, Italy) \emph{(\bibinfo{series}{CIKM '18})}.
  \bibinfo{publisher}{Association for Computing Machinery},
  \bibinfo{address}{New York, NY, USA}, \bibinfo{pages}{2165–2173}.
\newblock
\showISBNx{9781450360142}
\urldef\tempurl%
\url{https://doi.org/10.1145/3269206.3272018}
\showDOI{\tempurl}


\bibitem[\protect\citeauthoryear{Xu, Poole, Miller, Eiriksdottir, Kestranek,
  Catrambone, and Mynatt}{Xu et~al\mbox{.}}{2012}]%
        {xu2012not}
\bibfield{author}{\bibinfo{person}{Yan Xu}, \bibinfo{person}{Erika~Shehan
  Poole}, \bibinfo{person}{Andrew~D Miller}, \bibinfo{person}{Elsa
  Eiriksdottir}, \bibinfo{person}{Dan Kestranek}, \bibinfo{person}{Richard
  Catrambone}, {and} \bibinfo{person}{Elizabeth~D Mynatt}.}
  \bibinfo{year}{2012}\natexlab{}.
\newblock \showarticletitle{This is not a one-horse race: understanding player
  types in multiplayer pervasive health games for youth}. In
  \bibinfo{booktitle}{\emph{Proceedings of the ACM 2012 conference on computer
  supported cooperative work}}. \bibinfo{publisher}{Association for Computing
  Machinery}, \bibinfo{address}{New York, NY, USA}, \bibinfo{pages}{843--852}.
\newblock


\bibitem[\protect\citeauthoryear{Zhang and Hurley}{Zhang and Hurley}{2008}]%
        {zhang2008avoiding}
\bibfield{author}{\bibinfo{person}{Mi Zhang} {and} \bibinfo{person}{Neil
  Hurley}.} \bibinfo{year}{2008}\natexlab{}.
\newblock \showarticletitle{Avoiding monotony: improving the diversity of
  recommendation lists}. In \bibinfo{booktitle}{\emph{Proceedings of the 2008
  ACM conference on Recommender systems}}. \bibinfo{pages}{123--130}.
\newblock


\bibitem[\protect\citeauthoryear{Zhao, Grasmuck, and Martin}{Zhao
  et~al\mbox{.}}{2008}]%
        {zhao2008identity}
\bibfield{author}{\bibinfo{person}{Shanyang Zhao}, \bibinfo{person}{Sherri
  Grasmuck}, {and} \bibinfo{person}{Jason Martin}.}
  \bibinfo{year}{2008}\natexlab{}.
\newblock \showarticletitle{Identity construction on Facebook: Digital
  empowerment in anchored relationships}.
\newblock \bibinfo{journal}{\emph{Computers in human behavior}}
  \bibinfo{volume}{24}, \bibinfo{number}{5} (\bibinfo{year}{2008}),
  \bibinfo{pages}{1816--1836}.
\newblock


\bibitem[\protect\citeauthoryear{Zhou, Kuscsik, Liu, Medo, Wakeling, and
  Zhang}{Zhou et~al\mbox{.}}{2010}]%
        {zhou2010solving}
\bibfield{author}{\bibinfo{person}{Tao Zhou}, \bibinfo{person}{Zolt{\'a}n
  Kuscsik}, \bibinfo{person}{Jian-Guo Liu}, \bibinfo{person}{Mat{\'u}{\v{s}}
  Medo}, \bibinfo{person}{Joseph~Rushton Wakeling}, {and}
  \bibinfo{person}{Yi-Cheng Zhang}.} \bibinfo{year}{2010}\natexlab{}.
\newblock \showarticletitle{Solving the apparent diversity-accuracy dilemma of
  recommender systems}.
\newblock \bibinfo{journal}{\emph{Proceedings of the National Academy of
  Sciences}} \bibinfo{volume}{107}, \bibinfo{number}{10}
  (\bibinfo{year}{2010}), \bibinfo{pages}{4511--4515}.
\newblock


\end{thebibliography}

\newpage
\appendix

\begin{figure*}[t]
    \centering
    \includegraphics[width=0.8\textwidth]{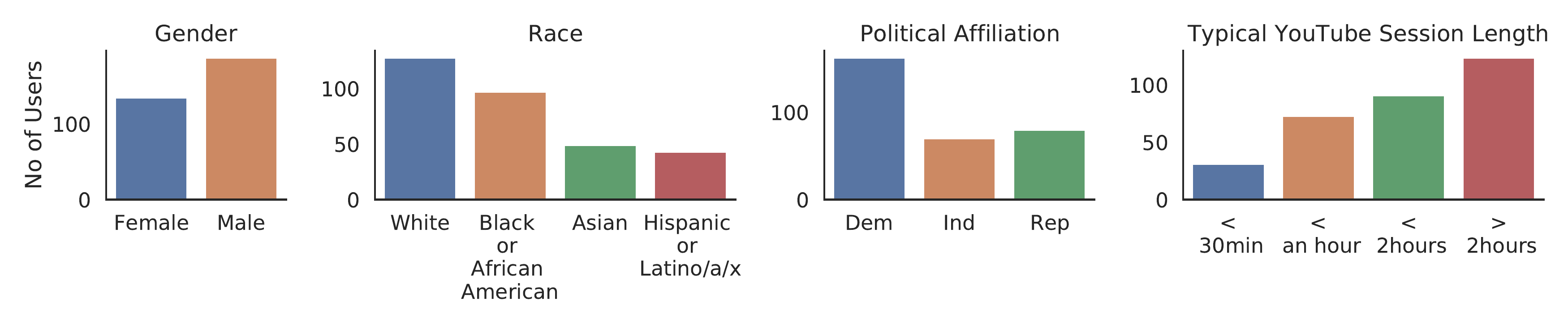}
    \vspace{-8pt}
    \caption{Demography of the participants who passed screening criteria and signed up for the study.}
    \label{fig:demo_all}
    \Description[User demography who passed screening test]{There are four bar charts in this figure, respectively for gender, race, political affiliation and typical YouTube session length. Here, both gender-wise and race-wise, it is balanced.}
\end{figure*}

\section{Distribution of Users Who Passed Eligibility Criteria and Signed Up}\label{app:demo}

Total 318 users signed up for our study. Figure \ref{fig:demo_all} shows the demography of these users.

\section{Need for Reflection and Insight Questionnaire}\label{app:scale}
These questions have been taken from Halttu et. al.~\cite{Halttu2017}.

\subsection{Need for Self-Reflection}
\begin{itemize}
    \item I am not really interested in analyzing my behavior. (R)
    \item It is important for me to evaluate the things that I do.
    \item I am very interested in examining what I think about.
    \item It is important to me to try to understand what my feelings mean.
    \item I have a definite need to understand the way that my mind works.
    \item It is important for me to be able to understand how my thoughts arise.
\end{itemize}

\subsection{Insight}
\begin{itemize}
    \item I usually have a very clear idea about why I’ve behaved in a certain way.
    \item My behavior often puzzles me. (R)
    \item Thinking about my thoughts makes me more confused. (R)
    \item Often I find it difficult to make sense of the way I feel about things.(R)
    \item I usually know why I feel the way I do.
\end{itemize}

\section{Semi-Structured Interview Questions}\label{app-interview}

\begin{itemize}
    \item How do you typically use YouTube to find new content?
    \item How would you describe the benefits and limitations in the features provided by YouTube to find new content?
    \item Can you walk me through your recommendation feed? How did you use features [sharing video, setting persona, browsing \sys] in \sys?
    \item What were your impressions when you saw strangers’ profiles and recommended videos?
    \item Could you describe any profile/recommended videos from the strangers during the study that stood out or were memorable to you? Why?
    \item How would you compare strangers’ recommendations to your YouTube recommendation?
    \item Would you continue to use \sys{} after the study? If so, what would be the purpose/motivation? If not, why not?
    \item Among the features provided in \sys{} about choosing your persona, what was important for you to present about yourself to others?
    \item How would you describe the features offered by the tool to share/remove recommended videos from your feed?
    \item What concerns would you have regarding sharing your profile and recommendation?
    \item Is there anything else where the tool should be more transparent about?
    \item Other than using the plugin, was there anything that you had to do and you have not done regularly for  this study?
    \item What did you like about the tool? What did you not like about the tool? Do you have any suggested changes on the tool for us to improve it?
\end{itemize}

\end{document}